\documentclass[twocolumn,english,aps,prb]{revtex4}
\usepackage{color}
\usepackage{amsmath}
\usepackage{graphicx}
\usepackage{amssymb}

\makeatletter
\@ifundefined{textcolor}{}
{%
 \definecolor{BLACK}{gray}{0}
 \definecolor{WHITE}{gray}{1}
 \definecolor{RED}{rgb}{1,0,0}
 \definecolor{GREEN}{rgb}{0,1,0}
 \definecolor{BLUE}{rgb}{0,0,1}
 \definecolor{CYAN}{cmyk}{1,0,0,0}
 \definecolor{MAGENTA}{cmyk}{0,1,0,0}
 \definecolor{YELLOW}{cmyk}{0,0,1,0}
 }

\@ifundefined{definecolor}
 {\usepackage{color}}{}
\@ifundefined{definecolor}{\@ifundefined{definecolor}
 {\usepackage{color}}{}
}{}

\newcommand{\sgn}{\operatorname{sgn}}
\newcommand{\w}{\omega}

\newcommand{\ba}{\begin{eqnarray*}}
\newcommand{\ea}{\end{eqnarray*}}
\newcommand{\baa}{\begin{eqnarray}}
\newcommand{\eaa}{\end{eqnarray}}
\newcommand{\bea}{\begin{eqnarray}}
\newcommand{\eea}{\end{eqnarray}}
\newcommand{\be}{\begin{equation}}
\newcommand{\ee}{\end{equation}}

\newcommand{\sm}{SmB$_6$}
\newcommand{\nimp}{n_{\rm imp}}
\newcommand{\Nimp}{N_{\rm imp}}
\newcommand{\rqpi}{\rho_{\rm QPI}}
\newcommand{\dbulk}{\Delta_{\rm bulk}}

\newcommand{\bk}{\mathbf{k}}
\newcommand{\bq}{\mathbf{q}}
\newcommand{\bQ}{\mathbf{Q}}
\newcommand{\br}{\mathbf{r}}
\DeclareMathOperator{\Tr}{Tr}
\DeclareMathOperator{\I}{Im}
\DeclareMathOperator{\R}{Re}
\makeatother

\usepackage{babel}

\begin{document}

\title{
Kondo holes in topological Kondo insulators: \\
Spectral properties and surface quasiparticle interference
}

\author{Pier Paolo Baruselli}
\author{Matthias Vojta}
\affiliation{Institut f\"ur Theoretische Physik, Technische Universit\"at Dresden, 01062 Dresden, Germany}


\begin{abstract}
A fascinating type of symmetry-protected topological states of matter are topological
Kondo insulators, where insulating behavior arises from Kondo screening of localized
moments via conduction electrons, and non-trivial topology emerges from the structure of
the hybridization between the local-moment and conduction bands.
Here we study the physics of Kondo holes, i.e., missing local moments, in three-dimensional
topological Kondo insulators, using a self-consistent real-space mean-field theory. Such
Kondo holes quite generically induce in-gap states which, for Kondo holes at or near the
surface, hybridize with the topological surface state. In particular, we study the
surface-state quasiparticle interference (QPI) induced by a dilute concentration of surface Kondo
holes and compare this to QPI from conventional potential scatterers.
We treat both strong and weak topological-insulator phases and, for the latter, specifically discuss the contributions to QPI from inter-Dirac-cone scattering.
\end{abstract}

\date{\today}


\maketitle

\section{Introduction}

In the exciting field of topological insulators,\cite{tirev1,tirev2} topological Kondo
insulators (TKIs) play a particularly interesting role: in these strongly correlated
electron systems, theoretically proposed in Refs.~\onlinecite{tki1,tki2}, a topologically
non-trivial bandstructure emerges at low energies and temperatures from the Kondo
screening of $f$-electron local moments due to the specific form of the
hybridization between conduction and $f$ electrons. As with standard topological
insulators, TKIs can exist in both two and three space dimensions, in the latter case as
strong and weak topological insulators. TKIs display helical low-energy surface states
which are expected to be heavy in the heavy-fermion sense, i.e., with a strong mass
renormalization and a small quasiparticle weight.

The material {\sm} was proposed to be a three-dimensional (3D) TKI,\cite{tki1,tki2,lu_smb6_gutz} and a number of recent experiments appear to support this hypothesis: transport studies have been interpreted in terms of quantized surface transport,\cite{wolgast_smb6} quantum oscillation measurements indicate the presence of a two-dimensional Dirac state,\cite{lixiang_smb6} and results from photoemission measurements\cite{neupane_smb6,mesot_smb6} and scanning tunneling spectroscopy (STS)\cite{hoffman_smb6} appear consistent with this assertion.
However, to date the topological nature of the surface states of {\sm} has not been unambiguously verified. Moreover, it has been suggested\cite{sawatzky_smb6} that the observed surface metallicity is polarity-driven, raising questions about the proper interpretation of the experimental data. This calls for more detailed studies of the surface-state physics of {\sm} and other candidate TKI materials.

A powerful probe of the surface electronic structure is Fourier-transform scanning tunneling spectroscopy\cite{eigler,davis_science} (FTSTS), applied in recent years, e.g., to both cuprate and iron-pnictide superconductors. Within FTSTS energy-dependent spatial variations of the local density of states (LDOS) are analyzed in terms of quasiparticle interference (QPI), i.e., elastic quasiparticle scattering processes due to impurities.
Such experiments were performed on the topological insulators Bi$_{1-x}$Sb$_x$ and Bi$_2$Te$_3$, and the results were found to be consistent with a suppression of backscattering, $\bk \leftrightarrow (-\bk)$, due to the spin-momentum locking of the helical surface state.\cite{yazdani,kapitulnik,xue,xia,yazdani11}

Theoretically, impurity scattering and QPI on the surface of 3D topological insulators have been studied for lattice models,\cite{wang_ti_imp} and within effective surface theories for non-magnetic,\cite{guo_franz_sti,lee_qpi} magnetic,\cite{guo_franz_sti,zhou_tmatrix_sti} and Kondo \cite{mitchell_kondo_ti,orignac_kondo_ti} impurities. In these works the surface electrons were assumed to be non-interacting, such that the interplay of impurity and strong-correlation effects, expected to be important for TKIs, has not been covered.

In this paper we aim at closing this gap, by studying the physics of local defects in Anderson lattice models of TKIs. In particular we will focus at local-moment vacancies, so-called Kondo holes.\cite{schlottmann_1,schlottmann_2} QPI from Kondo holes has been considered before\cite{morr_kh,davis_kh} for conventional heavy-fermion metals and has been found to be particularly revealing due to the interplay of defect and Kondo physics. Kondo holes on the surface of TKIs promise to be interesting also because they represent strong scatterers for which the simplest arguments of topological protection no longer apply.\cite{balatsky_imp_12,fehske}
Here we will employ a fully self-consistent mean-field description of the Kondo insulator, taking into account the local modification of Kondo screening by defects. Applying this methodology to both the weak topological insulator (WTI) and strong topological insulator (STI) phases, we will calculate the electronic structure and the surface QPI patterns for dilute surface Kondo holes as well as for other types of impurities. We will also present selected results for a finite concentration of Kondo holes.

%

\subsection{Summary of results}

Our main results can be summarized as follows.
Kondo holes on the surface of TKIs tend to create localized states, which hybridize with surface states. This gives rise to distinct features in the LDOS in the immediate vicinity of the hole, with an energy dependence mainly determined by the degree of particle--hole symmetry breaking: in the present model, the WTI phase occurs closer to the Kondo limit and is less particle--hole asymmetric, such that a strong in-gap resonance occurs. In the STI phase, particle--hole symmetry is strongly broken, and the hole-induced weight in the LDOS is shifted to elevated energies.

As expected, QPI patterns closely reflect the dispersion of surface states and are distinctly different for STI and WTI phases, with one and two surface Dirac cones, respectively. In the STI case the QPI signal close to the Dirac point is weak and weakly momentum-dependent, due to forbidden backscattering within a single Dirac cone. In contrast, the WTI displays a strong and strongly peaked QPI signal arising from intercone scattering.

To gain analytical insights into intercone scattering, we have extended the continuum Born-limit calculation of Ref.~\onlinecite{guo_franz_sti} to two Dirac cones for non-magnetic impurities, and we also sketch the extension for magnetic ones.

Our comparison of different types of impurities reveals surprisingly strong differences in the resulting QPI patterns, arising from (i) extended scattering regions (as compared to point-like defects) for Kondo holes due to a modification of the Kondo effect in the hole's vicinity and (ii) real parts of Greens functions entering the QPI signal invalidating the naive joint-density-of-states picture. In turn, this implies that experimental QPI results, in connection with careful modelling, can be used to determine the nature of the underlying scatterers.

For a finite concentration of Kondo holes, we find the expected disorder-induced broadening of the surface states. In the WTI phase, the low-energy resonances hybridize to yield an impurity-induced band.

On a technical level, we note that the Kondo effect is strongly modified both at the surface and near vacancies as compared to the bulk of the system, rendering fully self-consistent calculations necessary for a reasonably accurate description of QPI.

\subsection{Outline}

The body of the paper is organized as follows.
In Section~\ref{sec:model}  we briefly describe the model for TKIs and the type of impurities we studied.
Section~\ref{sec:mf} summarizes the slave-boson mean-field treatment for the translationally invariant case; its modifications for systems with surfaces and/or impurities are described in Section~\ref{sec:rmf}. In particular, we discuss how to efficiently calculate propagators for the case of isolated Kondo holes with fully self-consistent mean-field parameters.
Numerical results are shown in the remainder of the paper, starting with the clean system in Section~\ref{sec:clean}. The main body of results is given in Section~\ref{sec:oneimp} for isolated impurities, covering the impurity-induced density of states and the QPI patterns. Finally, Section~\ref{sec:manyimp} presents single-particle spectra for disordered systems with finite concentration of surface Kondo holes.
In Section~\ref{sec:concl} we present the conclusions of our work.


\section{Modelling}
\label{sec:model}

\subsection{Anderson lattice model for topological Kondo insulator}

Our work utilizes a tight-binding lattice model for a three-dimensional topological Kondo insulator. Following Refs. \onlinecite{tki1, tki2}, we consider a periodic Anderson lattice on a simple cubic (more precisely, tetragonal) lattice, with the Hamiltonian
\begin{eqnarray}
\label{H_r}
H_0&=&-t_c \sum_{\langle ij \rangle \sigma} (c_{i\sigma}^\dagger c_{j\sigma}+h.c.)+\nonumber\\
&+&\epsilon_{f} \sum_{i\alpha}{f}_{i\alpha}^\dagger {f}_{i\alpha} - t_f\sum_{\langle ij \rangle\alpha} ({f}_{i\alpha}^\dagger {f}_{j\alpha}+h.c.)+\nonumber\\
&+& V \sum_{\langle ij \rangle\sigma\alpha}(\Phi_{i\sigma j\alpha}c_{i\sigma}^\dagger {f}_{j\alpha}+h.c.)+\nonumber\\
&+& U \sum_{i} f_{i+}^\dagger f_{i+} f_{i-}^\dagger f_{i-},
\end{eqnarray}
in standard notation. The model entails two doubly degenerate orbitals per site, labelled $c$ for conduction electrons and $f$ for localized $f$-shell electrons, respectively. The index $\sigma=\uparrow,\downarrow$ denotes the spin of $c$ electrons, while the index $\alpha=+,-$ corresponds to the pseudo-spin of the $f$ electrons. Both hopping and hybridization terms are assumed to be non-zero for pairs $\langle ij\rangle$ of nearest-neighbor sites only, with $t_c>0$, $t_f<0$, $V>0$. Note that non-zero $f$ hopping is required to yield a finite band gap within the slave-boson approximation described below.
We will employ $t_c=1$ as energy unit unless otherwise noted.

The operators ${f}_{i\alpha}$ describe the lowest Kramers doublet of the $f$ electrons once spin-orbit interaction and crystal-field splitting are taken into account, for details see Ref. \onlinecite{tki2}. Here we choose a situation corresponding to tetragonal symmetry, with the lowest doublet being $\Gamma_{8(2)}$ constituted by states with $J=\frac{5}{2}$ and $\left|J_z\right|=\frac{1}{2}$ ($\eta=\arctan(2/\sqrt{3})$):
\begin{eqnarray}
|+\rangle&\equiv|J_z=+\frac{1}{2}\rangle= &\cos\eta|0\uparrow\rangle-\sin\eta|+1\downarrow\rangle,\label{+}\\
|-\rangle&\equiv|J_z=-\frac{1}{2}\rangle= &\sin\eta|-1\uparrow\rangle-\cos\eta|0\downarrow\rangle,\label{-}
\end{eqnarray}
where $\uparrow$, $\downarrow$ here denotes the spin degree of freedom for $f$ electrons, while $-1$, $0$, $+1$ is $m$, the azimuthal quantum number
for angular momentum $L=3$.
Experimentally, this may be realized, e.g., in tetragonal Ce compounds with dominant $f^1$ configuration, provided that the $f$ electron resides in the chosen doublet.

We note that other ground-state doublets may be considered; however, to our knowledge, this is the only choice compatible with tetragonal symmetry which grants bulk-insulating behavior in 3D. For example, the $\Gamma_{8(1)}$ doublet used in Ref.~\onlinecite{assaad_tki_dmft} and in Ref.~\onlinecite{tki_kim} (together with $\Gamma_{8(2)}$) generates an insulator in 2D, but only a semimetal in 3D, as it provides no hybridization along the third direction.
An appealing alternative, more tailored towards \sm, would be to consider a cubic environment.\cite{tki_cubic} However, this implies a degeneracy of the $f$ multiplet to be 4 rather than 2, thus significantly complicating the theoretical analysis. We expect that most of the features we find are generic, i.e., would also apply to the cubic case, but we leave a more detailed study of the latter for future work.


The non-trivial topological  behavior of the model is encoded in the $c-f$ hybridization form factor\cite{tki_kim} $\Phi_{i\sigma j\alpha}$
between a $c$ electron at site $i$ with spin $\sigma$, and an $f$ electron at site $j$ with pseudo-spin $\alpha$.
It is defined (up to a constant factor) by the overlap between their wavefunctions:
\begin{eqnarray}
\Phi_{i\sigma j \alpha}&\equiv& \langle i \sigma | j \alpha  \rangle= \langle \br_i-\br_j \sigma | 0  \alpha  \rangle=\nonumber\\
&=&\sum_{m\sigma'} A^\alpha_{m \sigma'} \langle \br_i-\br_j \sigma|0 m\sigma'\rangle=\nonumber\\
&=&\sum_m A^\alpha_{m \sigma} Y_m^3(\Omega_{\br_i-\br_j}),
\end{eqnarray}
which holds if $\br_i$ and $\br_j$ are nearest neighbors, otherwise $\Phi_{i\sigma j \alpha}$ is assumed to be zero;
coefficients $A^\alpha_{m \sigma}$ are taken from Eqs. \eqref{+}, \eqref{-}, and $Y_m^3$ are the spherical harmonics for $L=3$, and azimuthal quantum number $m$.
From $Y_{-1}^3=-Y_1^{3*}$ we get
\begin{eqnarray}
\Phi_{i\sigma j \alpha}&=&(\Phi_{ij})_{\sigma \alpha}=\nonumber\\
&=&\left(\!
\begin{array}{cc}
 +\cos\eta Y_0^3(\Omega_{\br_i-\br_j}) &-\sin\eta Y_1^{3*}(\Omega_{\br_i-\br_j}) \\
-\sin\eta Y_1^3(\Omega_{\br_i-\br_j}) &-\cos\eta Y_0^3(\Omega_{\br_i-\br_j})
\end{array}\!\right).
\end{eqnarray}

Using the explicit expressions of the spherical harmonics
\begin{eqnarray}
Y_{1}^3(\theta,\phi)&=&-\frac{1}{8}\sqrt\frac{21}{\pi}e^{i\phi}\sin\theta(5\cos^2\theta-1),\\
Y_{0}^3(\theta,\phi)&=&\frac{1}{4}	\sqrt\frac{7}{\pi}(5\cos^3\theta-3\cos\theta),
\end{eqnarray}
we find: 
\begin{eqnarray}
\Phi_{i\sigma 0 \alpha}^{+x}=-\frac{1}{4}\sqrt\frac{3}{\pi}\sigma_x, \hspace{10pt}\Phi_{i\sigma 0 \alpha}^{-x}=+\frac{1}{4}\sqrt\frac{3}{\pi}\sigma_x, \\
\Phi_{i\sigma 0 \alpha}^{+y}=-\frac{1}{4}\sqrt\frac{3}{\pi}\sigma_y, \hspace{10pt}\Phi_{i\sigma 0 \alpha}^{-y}=+\frac{1}{4}\sqrt\frac{3}{\pi}\sigma_y,\\
\Phi_{i\sigma 0 \alpha}^{+z}=+\frac{1}{2}\sqrt\frac{3}{\pi}\sigma_z, \hspace{10pt}\Phi_{i\sigma 0 \alpha}^{-z}=-\frac{1}{2}\sqrt\frac{3}{\pi}\sigma_z,
\end{eqnarray}
or, absorbing the $\sqrt{3/\pi}/4$ factor in $V$:
\begin{equation}
\Phi_{i\sigma j\alpha}=\begin{cases}\label{phi_ij}
-(\sigma_x)_{\sigma\alpha}\sgn(x_i-x_j), &\mbox{$\langle{i,j}\rangle$ n.n. along $x$,}\\
-(\sigma_y)_{\sigma\alpha}\sgn(y_i-y_j), &\mbox{$\langle{i,j}\rangle$ n.n. along $y$,}\\
2(\sigma_z)_{\sigma\alpha}\sgn(z_i-z_j),&\mbox{$\langle{i,j}\rangle$ n.n. along $z$,}\\
0, & \mbox{otherwise.}
                      \end{cases}
\end{equation}
Note that $\Phi_{i\sigma j\alpha}=-\Phi_{j\sigma i\alpha}$: this is equivalent to $\Phi_{\bk\sigma\alpha}=-\Phi_{-\bk\sigma\alpha}$,
which is needed to ensure the non-trivial topological behavior of the model.\cite {tki1,tki2}

The model described by Eqs.~\eqref{H_r} and \eqref{phi_ij} can realize a multitude of different phases, depending on the band filling and the values of $U$, $V$, $t_f$, and $\epsilon_f$. A mean-field phase diagram will be discussed in Section~\ref{sec:mf} below.

\subsection{Kondo holes and other impurities}

Most generally, we will consider random potentials on both the $c$ and $f$ orbitals, described by a disorder Hamiltonian
\begin{eqnarray}
H_{\rm dis}&=&\sum_{i\sigma} \Delta\epsilon_{ci} c_{i\sigma}^\dagger c_{i\sigma}
+\sum_{i\alpha} \Delta\epsilon_{fi} f_{i\alpha}^\dagger f_{i\alpha}.
\end{eqnarray}
The full Hamiltonian is then given by $H=H_0+H_{\rm dis}$, where we can define site-dependent potentials according to $\epsilon_{ci}=\Delta\epsilon_{ci}$ and $\epsilon_{fi}=\Delta\epsilon_{fi}+\epsilon_f$.

Kondo holes represent sites $i$ with missing $f$-orbital degrees of freedom (i.e. non-magnetic ions); they are modelled by $\Delta\epsilon_{fi}\to\infty$ and $\Delta\epsilon_{ci}=0$
(in practice we use $\Delta\epsilon_{fi}=100 t_c$).
We will also consider weak scatterers in either the $c$ or the $f$ band, described by small non-zero
$\Delta\epsilon_{ci}$ or $\Delta\epsilon_{fi}$, respectively.

A large part of the paper is devoted to the study of isolated impurities, where only a single site $i$ has non-vanishing $\Delta\epsilon_{ci}$ or $\Delta\epsilon_{fi}$, but Section~\ref{sec:manyimp} will also consider the case of a finite number $\Nimp$ (or finite concentration $\nimp$) of defect sites with non-vanishing $\Delta\epsilon_{ci}$ or $\Delta\epsilon_{fi}$.



\section{Mean-field theory: Translation-invariant case}
\label{sec:mf}

When translation symmetry holds,
the one-body part of Eq. \eqref{H_r} can be Fourier-transformed to yield:
\begin{eqnarray}\label{H_k}
H_0&=& -t_c\sum_{\bk\sigma} F_\bk c_{\bk\sigma}^\dagger c_{\bk\sigma}+\sum_{\bk\alpha}(\epsilon_f-t_f F_\bk){f}_{\bk\alpha}^\dagger {f}_{\bk\alpha}+\nonumber\\
&&+ \sum_{\bk\sigma\alpha}V(\Phi_{\bk\sigma\alpha}c_{\bk\sigma}^\dagger {f}_{\bk\alpha}+h.c.)+\nonumber\\
&&+\sum_{i} U f_{i+}^\dagger f_{i+} f_{i-}^\dagger f_{i-},
\end{eqnarray}
where  $\bk\equiv(k_x,k_y,k_z)$ is a momentum from the first Brillouin zone (BZ) ,$-\pi\leq k_x,k_y,k_z<\pi$, further
\begin{equation}
F_\bk=2(\cos k_x +\cos k_y +\cos k_z),\\
\end{equation}
and the $c-f$ hybridization $\Phi_{\bk\sigma\alpha}$ is the Fourier transform of Eq. \eqref{phi_ij}:
\begin{eqnarray}
\hat\Phi_\bk&=&d(\bk)\cdot\hat{\boldsymbol{\sigma}},\\
d(\bk)&=&(-2 i \sin k_x,-2 i\sin k_y,4 i\sin k_z),\\
\hat{\boldsymbol{\sigma}}&=&(\sigma_x,\sigma_y,\sigma_z).
\end{eqnarray}

\subsection{Slave-boson approximation}

To deal with the Coulomb repulsion $U$ in Eq.~\eqref{H_k}, we employ the slave-boson mean-field approximation,\cite{read_newns_slavebosons_1,read_newns_slavebosons, hewson} which is known to be reliable at low temperatures below the Kondo temperature.

The approximation is based on taking the limit $U\rightarrow \infty$, i.e., excluding doubly occupied $f$ orbitals. The remaining states of the local $f$ Hilbert space are represented by auxiliary particles, $b_i$ and $\tilde{f}_{i\alpha}$, for empty and singly occupied $f$ orbitals, respectively, such that ${f}_{i\alpha}= b_i^\dagger \tilde{f}_{i\alpha}$. The Hilbert space is constrained by $b_i^\dagger b_i +\sum_\alpha \tilde{f}^\dagger_{i\alpha}\tilde{f}_{i\alpha}=1$.
It is convenient to choose $b_i$ bosonic and $\tilde{f}_{i\alpha}$ fermionic, and to employ a saddle-point approximation $b_i \rightarrow b=\langle b_i \rangle$. With fluctuations of $b_i$ frozen, the above constraint is imposed in a mean-field fashion using a Lagrange multiplier $\lambda$.
$H_0$ takes the bilinear form:
\begin{eqnarray}	
H_0^{\rm MF}&=&-t_c\sum_{\bk\sigma} F_\bk c_{\bk\sigma}^\dagger c_{\bk\sigma}+\sum_{\bk\alpha}(\epsilon_f-t_f F_\bk b^2)\tilde{f}_{\bk\alpha}^\dagger \tilde{f}_{\bk\alpha}+\nonumber\\
&&+ \sum_{\bk\sigma\alpha}bV(\Phi_{\bk\sigma\alpha}c_{\bk\sigma}^\dagger \tilde{f}_{\bk\alpha}+h.c.)-\nonumber\\
&&-\mu\left[\sum_{\bk\sigma} c_{\bk\sigma}^\dagger c_{\bk\sigma}+\sum_{\bk\alpha} \tilde{f}_{\bk\alpha}^\dagger \tilde{f}_{\bk\alpha}-N_e \right]-\nonumber\\
&&-\lambda\left[\sum_{\bk\alpha}{\tilde{f}_{\bk\alpha}^\dagger \tilde{f}_{\bk\alpha}}+N_s(b^2-1)\right],\label{F_k}
\end{eqnarray}
with $N_e$ the total number of electrons and $N_s$ the number of lattice sites. We have introduced the chemical potential $\mu$ as the Lagrange multiplier enforcing the average electron number to be $N_e$, with the Kondo insulator reached at $N_e=2N_s$.

Minimization of saddle-point free energy leads to the self-consistency equations \cite{read_newns_slavebosons_1,read_newns_mixedval,read_newns_review,tki_kim}
\begin{eqnarray}
\label{scf_mu}
N_e &=&\sum_{\bk\sigma} \langle c_{\bk\sigma}^\dagger c_{\bk\sigma}\rangle+\sum_{\bk\alpha} \langle \tilde{f}_{\bk\alpha}^\dagger \tilde{f}_{\bk\alpha}\rangle,\\
\label{scf_b}
0 &=& 2b\left(\frac{1}{N_s}\sum_{\bk\alpha}t_f F_\bk\langle \tilde{f}_{\bk\alpha}^\dagger \tilde{f}_{\bk\alpha}\rangle-\lambda\right)\nonumber+\\
&&+\frac{V}{N_s}\sum_{\bk\sigma\alpha}\left( \Phi_{\bk\sigma\alpha}\langle c_{\bk\sigma}^\dagger \tilde{f}_{\bk\alpha}\rangle+h.c.\right),\\
\label{scf_lambda}
1 &=& b^2+\frac{1}{N_s}\sum_{\bk\alpha}{\langle \tilde{f}_{\bk\alpha}^\dagger \tilde{f}_{\bk\alpha}\rangle},
\end{eqnarray}
which determine the parameters $b$, $\lambda$, and $\mu$.

The diagonalization of $H_0^{\rm MF}$ yields single-particle energies $e_{j\bk}$ and corresponding eigenvectors $w_{j\bk}$, with quantum numbers $\bk$ and $j=1,\ldots,4$. The expectation value $\langle \hat{O}_\bk \rangle$ of a momentum-diagonal single-particle operator $\hat{O}_\bk$
is then given by
\begin{equation}
\langle \hat{O}_\bk \rangle=\sum_{j=1}^4 \langle w_{j\bk}|\hat{O}_\bk| w_{j\bk}\rangle n_F(e_{j\bk}-\mu),
\end{equation}
where $n_F(\omega)=(e^{\omega/T} +1)^{-1}$ is the Fermi-Dirac distribution function and $T$ the temperature. Most of our calculations below are intended to be for $T=0$; practically we used $T=0.01$ to avoid discretization errors.

\subsection{Mean-field phases}

Within the slave-boson approximation the model in Eqs.~(\ref{H_r},\ref{phi_ij}) has been shown \cite{tki_kim} to have four different phases as a function of its parameters. For small $V$, one encounters a decoupled phase with $b=0$ which may be classified as a fractionalized Fermi liquid\cite{flst1} (FL$^\ast$), i.e., an orbital-selective Mott state. Upon increasing $V$, transitions occur to a WTI phase with topological indexes $(0;111)$, a STI phase with
$(1;000)$, and finally a trivial band-insulating (BI) phase with $(0;000)$, see also Fig.~\ref{fig_phase_diag} below.
This is in contrast with the standard Doniach model \cite{doniach} with on-site hybridisation $V\sum_\sigma(c_{i\sigma}^\dagger f_{i\sigma}+h.c.)$, which only shows a transition from the decoupled to the trivial insulating phase.
We note that more complicated mean-field phase diagrams can arise\cite{sigrist_tki} when introducing second- and third-nearest neighbor hopping into $H_0$, but no qualitatively new phases appear.
Beyond the present mean-field theory, antiferromagnetism can be expected for small $V$, but the phases at larger $V$ are likely robust.

As shown in Appendix \ref{app_sitte}, the mean-field Hamiltonian $H_0^{\rm MF}$ is equivalent to the common cubic-lattice four-band model\cite{qi2008,sitte_ti} used in the topological-insulator literature. However, in the presence of boundaries and impurities, the physics of the Kondo-insulator model is richer due to the additional self-consistency conditions.


\section{Real-space mean-field theory}
\label{sec:rmf}

In situations without full translation symmetry the local mean-field parameters $b$ and $\lambda$ become site-dependent, which requires to formulate the mean-field theory in real space. We shall assume the bare hopping matrix elements $t_c$, $t_f$, $V$ to be position-independent as in Eq. \eqref{H_r}, but we treat the case with arbitrary on-site energies in $H=H_0+H_{\rm dis}$. Then
\begin{eqnarray}\label{F_r}
H^{\rm MF}&=&\sum_{i\sigma}\epsilon_{ci} c_{i\sigma}^\dagger c_{i\sigma} - t_c \sum_{\langle ij \rangle \sigma} (c_{i\sigma}^\dagger c_{j\sigma}+h.c.)+\nonumber\\
&&+\sum_{i\alpha}\epsilon_{fi} \tilde{f}_{i\alpha}^\dagger \tilde{f}_{i\alpha} - t_f\sum_{\langle ij \rangle\alpha} b_i b_j (\tilde{f}_{i\alpha}^\dagger \tilde{f}_{j\alpha}+h.c.)+\nonumber\\
&&+ \sum_{\langle ij \rangle\sigma\alpha}V(b_j\Phi_{i\sigma j\alpha}c_{i\sigma}^\dagger \tilde{f}_{j\alpha}+h.c.)-\nonumber\\
&&-\mu\left[\sum_{i\sigma} c_{i\sigma}^\dagger c_{i\sigma}+\sum_{i\alpha} \tilde{f}_{i\alpha}^\dagger \tilde{f}_{i\alpha}-N_e \right]-\nonumber\\
&&-\sum_i \lambda_i\left(\sum_\alpha \tilde{f}_{i\alpha}^\dagger \tilde{f}_{i\alpha}+b_i^2-1\right).
\end{eqnarray}
The local mean-field equations for $b_i$ and $\lambda_i$ read:
\begin{eqnarray}
0&=&\sum_{\langle j_i \rangle \sigma\alpha} V\left( \Phi_{j\sigma i\alpha}\langle c_{j\sigma}^\dagger \tilde{f}_{i\alpha}\rangle+h.c.\right)-2\lambda_i b_i+\nonumber\\
&&+ \sum_{\langle j_i \rangle} t_f b_j \left(\sum_\alpha  \langle \tilde{f}_{i\alpha}^\dagger \tilde{f}_{j\alpha}\rangle+h.c. \right), \label{scf_bi}\\
1&=& b_i^2+\sum_\alpha \langle \tilde{f}_{i\alpha}^\dagger \tilde{f}_{i\alpha}\rangle, \label{scf_lambdai}
\end{eqnarray}
where $\langle j_i \rangle$ denotes a nearest neighbor of site $i$.
Practically, Eqs.~(\ref{F_r},\ref{scf_bi},\ref{scf_lambdai}) need to be solved numerically for finite-size systems.
For sites with Kondo holes, i.e. no $f$ degrees of freedom, we formally set $b_i=\lambda_i=0$ which correctly excludes hopping to these sites.

The chemical potential $\mu$ remains a global parameter controlling the electron concentration $n_e = N_e/N_s$. In the thermodynamic limit with $n_e$ fixed, $\mu$ will be insensitive to the existence of surfaces as well as to a finite number of impurities ($\Nimp/N_s\to 0$). This is no longer true for finite systems. In our simulations we will fix $\mu$ to its value determined for the translation-invariant case, which implies that $n_e$ can differ slightly from 2 in the presence of surfaces. The advantage of this protocol is to avoid complications arising from a size-dependent $\mu$.

To improve accuracy within a finite-size self-consistent calculation, we have employed supercells (equivalent to an average over twisted periodic boundary conditions). Unless noted otherwise, a $2\times 2$ supercell grid was used.


\subsection{Clean system in slab geometry}

Surface states are efficiently modelled in slab systems of size $N_s= n_x\times n_x\times N_z$, with open boundary conditions along $z$ and periodic boundary conditions along $x$ and $y$ directions. Then the in-plane momentum $\bk=(k_x,k_y)$ remains a good quantum number, and the mean-field parameters depend on $z$ only.
After a (partial) Fourier transform in the $xy$ plane, electron operators carry indices $z$ and $\bk$, and Eq. \eqref{F_r} can be written as
\begin{equation}
H^{\rm MF} = \sum_{\bk} H^{\rm MF}_\bk + \mu N_e + N_x^2 \sum_z \lambda_z (1-b_z^2)
\end{equation}
with
\begin{eqnarray}\label{F_kr}
H^{\rm MF}_{\bk}&=& -t_c \sum_{z\sigma}  F'_\bk c_{z\bk\sigma}^\dagger c_{z\bk\sigma} - t_c\sum_{\langle zz'\rangle \sigma}(c_{z\bk\sigma}^\dagger c_{z'\bk\sigma}+h.c.)+\nonumber\\
&&+\sum_{z\alpha}(\epsilon_{f}-t_f b_z^2 F_\bk) \tilde{f}_{z\bk\alpha}^\dagger \tilde{f}_{z\bk\alpha}+\nonumber\\
&&-t_f\sum_{\langle zz'\rangle\alpha} b_z b_{z'} (\tilde{f}_{z\bk\alpha}^\dagger \tilde{f}_{z'\bk\alpha}+h.c.)+\nonumber\\
&&+ V\sum_{z\sigma\alpha}(b_z\Phi'_{\bk\sigma \alpha}c_{z\bk\sigma}^\dagger \tilde{f}_{z\bk\alpha}+h.c.)-\nonumber\\
&&+ V\sum_{\langle zz'\rangle\sigma\alpha}(b_{z'}\Phi'_{zz'\sigma \alpha}c_{z\bk\sigma}^\dagger \tilde{f}_{z'\bk\alpha}+h.c.)-\nonumber\\
&&-\mu\sum_{z\sigma} c_{z\bk\sigma}^\dagger c_{z\bk\sigma} - \sum_{z\alpha} (\mu+\lambda_z) \tilde{f}_{z\bk\alpha}^\dagger \tilde{f}_{z\bk\alpha},
\end{eqnarray}
where $\langle zz'\rangle $ means $z'=z\pm1$, and with
\begin{eqnarray}
F'_\bk&=&2(\cos k_x + \cos k_y),\\
\Phi'_{\bk\sigma\alpha}&=&-2i\sin k_x (\sigma_x)_{\sigma\alpha}-2i\sin k_y (\sigma_y)_{\sigma\alpha},\\
\Phi'_{zz'\sigma\alpha}&=&2(\sigma_z)_{\sigma\alpha}\sgn(z-z').
\end{eqnarray}
The eigenstates of $H^{\rm MF}_{\bk}$, $|n\bk\rangle$, carry quantum numbers of in-plane momentum $\bk$ and band index $n=1,\ldots,4N_z$.

Self-consistency equation can be written down for $b_z$ and $\lambda_z$ in analogy to Eqs.~\eqref{scf_bi}, \eqref{scf_lambdai}, but are omitted here for space reasons. Practically, we have used $N_z=15$ which we found sufficient to suppress the finite-size gap arising from the coupling of the two surfaces.

\subsection{Single impurity: Embedding procedure}
\label{sec_embedding}

In the presence of impurities, the system becomes fully inhomogeneous such that the real-space equations (\ref{F_r},\ref{scf_bi},\ref{scf_lambdai}) need to be solved. This restricts the numerical calculation to relatively small system sizes ($N_s=10^3 \ldots 15^3$) which are insufficient to accurately study Friedel oscillation and quasiparticle interference.

\begin{figure}[!tbp]
\includegraphics[width=0.48\textwidth]{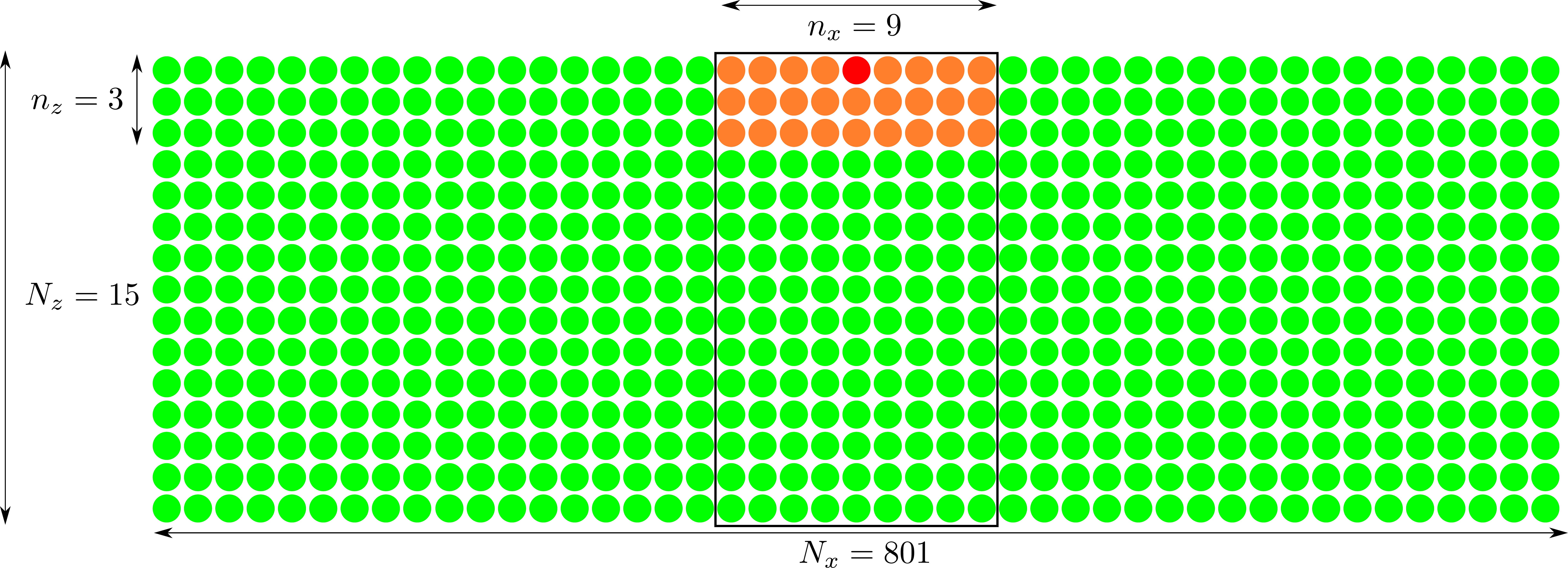}
\caption{Embedding procedure: a $n_x\times n_x\times n_z$ scattering region (red: impurity site, orange: sites with mean-field parameters different from the clean case)
is embedded in a larger $N_x\times N_x\times N_z$  region, where mean-field parameters are assumed to be unperturbed.
Black rectangle: $n_x\times n_x\times N_z$ region of full mean-field calculation in slab geometry.
}\label{fig_embedding}
\end{figure}

Notably, the problem can be simplified for isolated impurities using scattering-matrix techniques. The basic observation is that mean-field parameters parameters $b_i$ and $\lambda_i$ are locally perturbed by each impurity, but these perturbations decay on short length scales. Therefore, to a good approximation, electron scattering off each impurity can be described in terms of small-size scattering regions where $b_i$ and $\lambda_i$ deviate from their bulk values.

Specifically, for a single impurity in the surface layer we employ the following procedure, Fig.~\ref{fig_embedding}. We determine $b_i$ and $\lambda_i$ in a fully self-consistent inhomogeneous calculation for a small system of size $n_x\times n_x\times N_z$ in slab geometry. This scattering region is then embedded into a much larger $N_x\times N_x\times N_z$ system; for computational efficiency we further reduce the size of the scattering region to $n_z < N_z$ layers, because layers far from the impurity are only weakly perturbed.

Impurity-induced changes of electron propagators are then calculated using the T-matrix formalism:
\begin{equation}
\hat{G}(\omega)=\hat{G^0}(\omega)+\hat{G^0}(\omega)\hat{T}(\omega)\hat{G^0}(\omega),\label{gg0t}
\end{equation}
where the scattering matrix is determined as
\begin{equation}
\hat{T}(\omega)=\hat{V}\left(1-\hat{G^0}(\omega)\hat{V}\right)^{-1}.
\end{equation}
All matrices depend on indexes $x,y,z,s,a$ ($1\le x,y \le N_x$, $1\le z \le N_z$, $a=c/f$, $s=\uparrow,\downarrow$ if $a=c$, $s=+,-$ if $a=f$). The interaction matrix is given by
\begin{equation}
\hat{V}\equiv \hat{H}^{\rm MF}-\hat{H}_0^{\rm MF},
\end{equation}
where $\hat{H}^{\rm MF}$ is the mean-field Hamiltonian with defect and self-consistently determined $b_i$ and $\lambda_i$, and $\hat{H}_0^{\rm MF}$ the mean-field Hamiltonian of the clean slab.
$\hat{V}$, reflecting both the defect and its induced changes of mean-field parameters, is taken to be non-zero only in the small $n_x\times n_x\times n_z$ scattering region.

The Green's function of the impurity-free slab, $\hat{G^0}$, is diagonal in in-plane momentum $\bk=(k_x,k_y)$,
\begin{equation}
\label{gkslab}
\hat{G^0}_{zsa,z's'a'}(\omega,\bk)=\left(\hat{1}(\omega+\mu+i\delta)-\hat{H}^{\rm MF}_{\bk}\right)^{-1}_{zsa,z's'a'},
\end{equation}
with $\hat{H}^{\rm MF}_{\bk}$ from Eq.~\eqref{F_kr}, and $\delta$ is an artificial broadening parameter.
Fast Fourier transform is used to obtain $\hat{G^0}$ in real space.
Finally, the LDOS is computed through
\begin{equation}\label{eq_dos_t}
\rho(\omega,x,y,z)=-\frac{1}{\pi}\sum_{sa}\I \hat{G}_{xyzsa,xyzsa}(\omega)\,.
\end{equation}
For weak scatterers, the lowest-order Born approximation is sufficient, $\hat{T}\approx \hat{V}$.

For these scattering calculations we employed $N_x=801$ to achieve a high momentum resolution for QPI (see below), combined with a broadening $\delta=0.0006$ for the WTI and $\delta=0.002$ for the STI, to have a smooth energy-dependent DOS. $n_z=3$ was found to be sufficient to obtain converged results for the surface LDOS.

\subsection{Surface quasiparticle interference}

The QPI signal is obtained from the energy-dependent surface LDOS by Fourier transformation in the $xy$ plane:
\begin{equation}
\rqpi(\omega,\bk)=\frac{1}{N_x^2}\sum_{\bk}e^{i(k_x x+k_y y)} \Delta\rho(\omega,x,y,z=1),
\end{equation}
where only the impurity-induced change in the density of states is considered,
\begin{equation}
\label{eq_deltados_t}
\Delta\rho(\omega,x,y,z)=-\frac{1}{\pi}\sum_{sa}\I (\hat{G}-\hat{G}^0)_{xyzsa,xyzsa}(\omega);
\end{equation}
the homogeneous background would contribute a signal at $\bk=0$ only.
We note that $\rqpi$ is in general a complex quantity. However, in the case of a single impurity $\rho(x,y)$ is inversion-symmetric w.r.t. the impurity site, such that $\rqpi$ is real.

When relating the surface LDOS and $\rqpi$ to the signal in an actual STM measurement, complications arise from the fact that the differential tunneling current is not simply proportional to the LDOS: $c$ and $f$ signals are weighted differently, and an interference term is also present. \cite{madhavan98,kroha00,coleman09,morr10,balatsky_kondohole} Such corrections can be taken into account, but the required ratio of the different tunneling matrix elements into $c$ and $f$ orbitals is usually not known. Therefore we refrain from doing so; we anticipate that no qualitative changes to our conclusions would arise, although the energy dependence of features in the tunneling spectra may be modified.


\section{Results: Clean system}
\label{sec:clean}

We have investigated both the weak and strong topological-insulator phases of the model Eq.~\eqref{H_r}, with parameters chosen to obtain sufficiently large Kondo temperature and bulk gap, as to avoid finite-size effects, and a moderate surface-state Fermi velocity, as otherwise surface-state QPI is restricted to a tiny range in momentum space.


\subsection{Phase diagram}

\begin{figure}[!tbp]
\centering
\includegraphics[width=0.3\textwidth]{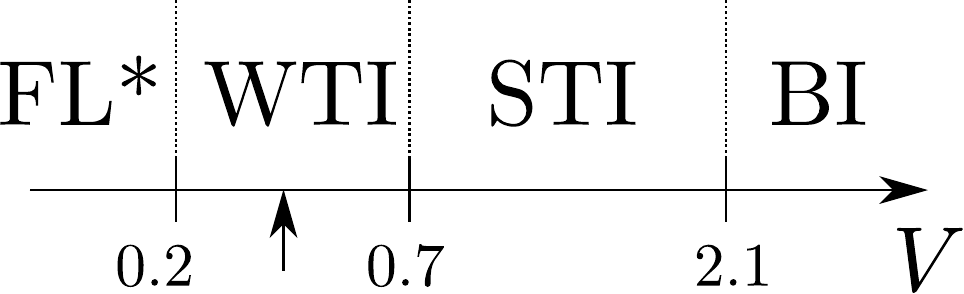}
\caption{Ground-state phase diagram of the model \eqref{H_r} in slave-boson mean-field approximation as function of $V$, for $t_c=1$, $t_f=-0.3$, $\epsilon_f=-1$.
FL$^\ast$= fractionalized Fermi liquid, WTI= weak topological insulator, STI=strong topological insulator, BI: trivial band insulator.
The arrow shows our parameter set for the WTI phase ($V=0.4$).
}\label{fig_phase_diag}
\end{figure}

Fig.~\ref{fig_phase_diag} shows a zero-temperature mean-field phase diagram, obtained from Eqs.~(\ref{F_k}-\ref{scf_lambda}), as function of the hybridization $V$ for the choice $t_c=1$, $t_f=-0.3$, $\epsilon_f=-1$. All phases except for FL$^\ast$ have $b\neq 0$; the WTI, STI, and BI phases have been detected by calculating the relevant $Z_2$ topological invariants of the mean-field bandstructure;\cite{tki2,Fu2007,fu_kane_ti_invsym} the transitions WTI$\leftrightarrow$STI
and STI$\leftrightarrow$BI can be detected via the closing of the bulk gap.

A general property of the model \eqref{H_r} is that the WTI phase is found deep in the Kondo regime, whereas the STI phase is realized in a regime of stronger valence fluctuations.


\subsection{Band structure and surface states}

Subjecting the system to open boundary conditions along $z$, metallic states appear on the two (001) surfaces in both the WTI and STI phases. For the WTI we found two Dirac cones at the two inequivalent momenta $(0,\pi)$ and $(\pi,0)$ of the surface Brillouin zone;
the STI has a single Dirac cone at $(\pi,\pi)$.

We note that, depending on parameters, the surface states may disperse such that constant-energy cuts near the Dirac energy display multiple band crossings in addition to those arising from the Dirac cones, which in turn complicates the QPI analysis. In our choice of parameters we tried to avoid such situations.

\begin{figure}[!tbp]
\includegraphics[width=0.48\textwidth]{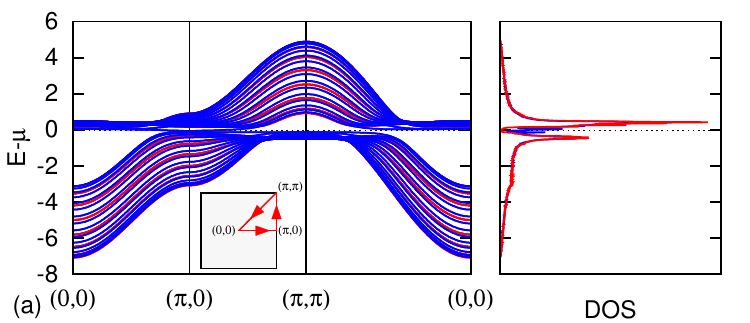}
\includegraphics[width=0.48\textwidth]{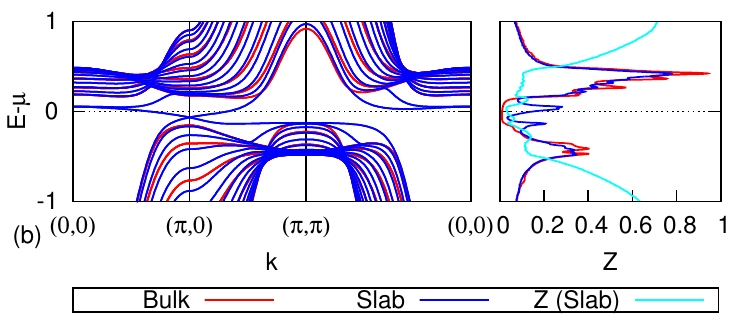}
\caption{Quasiparticle dispersion in the WTI phase, for $N_z=15$, $t_c=1$, $t_f=-0.3$, $V=0.4$, $\epsilon_f=-1$, for both periodic (``Bulk'') and open (``Slab'')
boundary conditions along $z$, together with associated density of states (an artificial broadening of 0.01 has been employed), (a) for the full energy range  and (b) in a  $2t_c$ window around the chemical potential .
For case (b), we also show the energy-dependent quasiparticle weight $Z$ for the slab case.
Momentum $k$ is taken along a path in the 2D Brillouin zone shown in the inset in case of panel (a).
There are two inequivalent $(\pi,0)$ and $(0,\pi)$ points, hence two inequivalent Dirac cones.
}\label{fig_bands_wti}
\end{figure}

\begin{figure}[!tbp]
\includegraphics[width=0.48\textwidth]{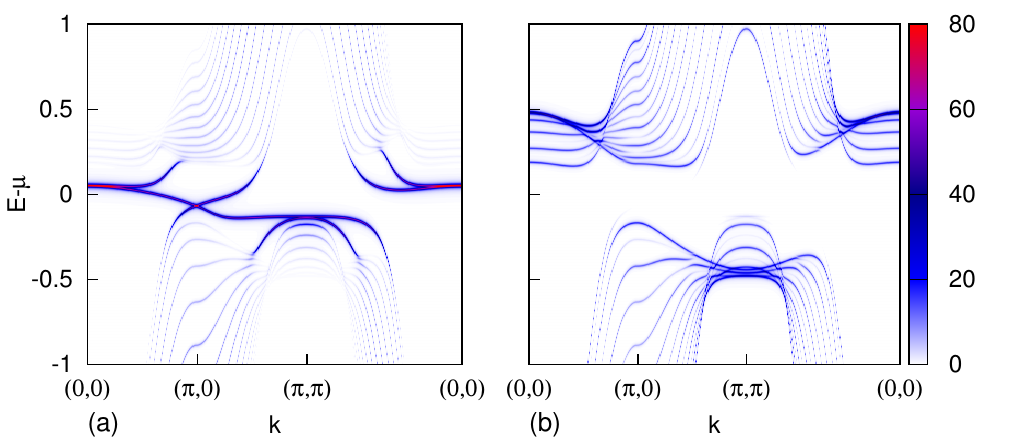}
\caption{Layer-resolved spectral intensity $A(\omega,\bk,z)$ \eqref{akslab} in the WTI phase (for the same parameters as Fig. \ref{fig_bands_wti}):
(a) on the surface ($z=1$) and, for comparison,
(b) in the bulk ($z=8$),
for the same path as in Fig. \ref{fig_bands_wti}.
An artificial broadening $\delta=0.005$ has been employed.
}
\label{fig_arpes_wti}
\end{figure}


\subsubsection{Weak topological insulator}
\label{sec:cleanweak}

Explicit results for the WTI phase have been obtained using $t_c=1$, $t_f=-0.3$, $V=0.4$, $\epsilon_f=-1$. The resulting mean-field band structure is shown in Fig.~\ref{fig_bands_wti} for both the periodic and slab cases.
The surface Dirac cones at momenta $(0,\pi)$ and $(\pi,0)$ are clearly visible, with the Dirac point at $E_{\textrm{Dirac}}-\mu=-0.066$; for our choice $N_z=15$ the cones display a tiny finite-size gap $\Delta_{fs}=7\times 10^{-4}$ due to the coupling between opposite surfaces.
We note that the choice of a relatively large value of $t_f$ is dictated by the necessity to obtain a sizeable bulk gap (which is zero when $t_f=0$), in order to have a sufficient energy window in which Dirac cones and the associated QPI can be studied. With our parameters, the bulk gap evaluates to  $\dbulk=0.28$.

Fig.~\ref{fig_arpes_wti} displays the the layer-resolved spectral intensity in the slab case, defined as
\begin{equation}
\label{akslab}
A(\omega,\bk,z) = -\frac{1}{\pi}\I\! \sum_{sa} \hat G^0_{zsa,zsa}(\omega,\bk)
\end{equation}
with $\hat G^0$ from Eq.~\eqref{gkslab}, illustrating the weight distribution for bulk and surface states.

All quasiparticle states $|n\bk\rangle$ are mixtures of $c$ and $f$ electrons. This may be quantified by the band- and momentum-dependent peak weight in the $c$-electron spectral function, usually dubbed quasiparticle weight, $Z_{n\bk} \equiv |\langle c|n\bk\rangle|^2$. We define an energy-dependent quasiparticle weight $Z(\epsilon)$ according to
\begin{equation}
Z(\omega)\equiv \frac{\rho_c(\omega)}{\rho_c(\omega)+\rho_f(\omega)}= \frac {\sum_{n\bk} {Z_{n\bk}}\tilde\delta(\omega-\epsilon_{n\bk}) }  { \sum_{n\bk} \tilde\delta(\omega-\epsilon_{n\bk}) },
\end{equation}
where $\rho_c(\epsilon)$ and $\rho_f(\epsilon)$ are, respectively, the $c$ and $f$ contributions to the total density of states, and $\tilde\delta(\omega)$ is a Lorentzian of width $\delta$.
As common for Kondo systems, $Z$ is small near the Fermi level, as shown in Fig. \ref{fig_bands_wti}(b). As a result the surface states are primarily of $f$ character, with $Z_{\rm Dirac} = 0.05$. We note that $Z$ does not directly correspond to the effective-mass ratio $m/m^\ast$ because the bare $f$ band is dispersive and hence the $c$-electron self-energy momentum-dependent.

The bulk Kondo temperature is estimated as $T_K=0.6$ from the location of the mean-field phase transition where $b$ becomes non-zero upon cooling; this transition is well-known to become a crossover upon including corrections beyond mean fields. Let us point out that surface effect on the mean-field parameters are sizeable: In the bulk we have $b=0.51$, $\lambda=-2.10$ whereas on the surface $b_{z=1}=0.28$ and $\lambda_{z=1}=-2.04$, i.e., Kondo screening is suppressed at the surface.
This strongly influences the Dirac-cone velocity: its value $v=0.08$ is much smaller than $v=0.21$ which would be obtained from a non-self-consistent slab calculation using bulk values of $b$ and $\lambda$. In other words, the interplay of Kondo and surface physics increases the mass of the surface quasiparticles -- an effect only captured by fully self-consistent calculations.

\begin{figure}[!tbp]
\includegraphics[width=0.48\textwidth]{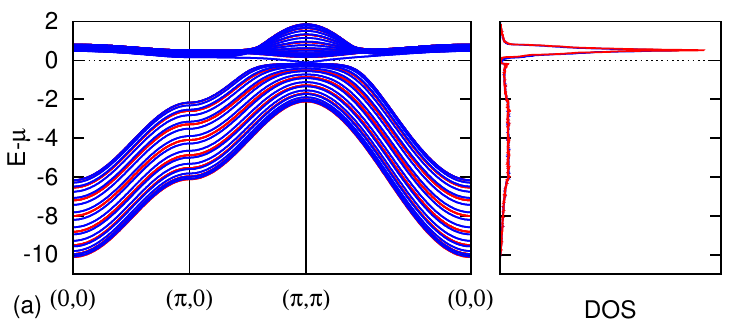}
\includegraphics[width=0.48\textwidth]{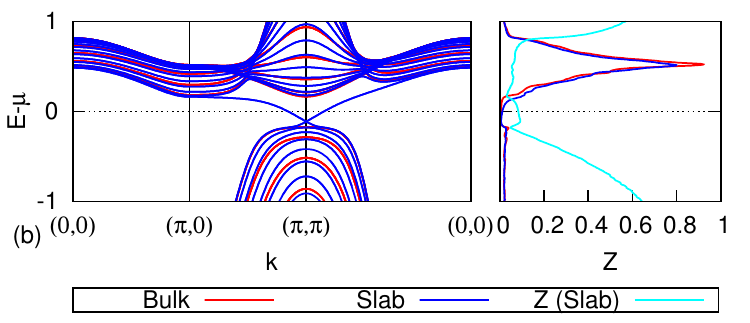}
\caption{Same as Fig. \ref{fig_bands_wti}, but now in the STI phase, for $N_z=15$, $t_c=1$, $t_f=-0.1$, $V=0.25$, $\epsilon_f=4$ (an artificial broadening of 0.01 has been employed),
for the full energy range (a) and in a $2t_c$ window around the chemical potential (b).
A single surface Dirac cone exists at $(\pi,\pi)$.
}\label{fig_bands_sti}
\end{figure}

\begin{figure}[!tbp]
\includegraphics[width=0.48\textwidth]{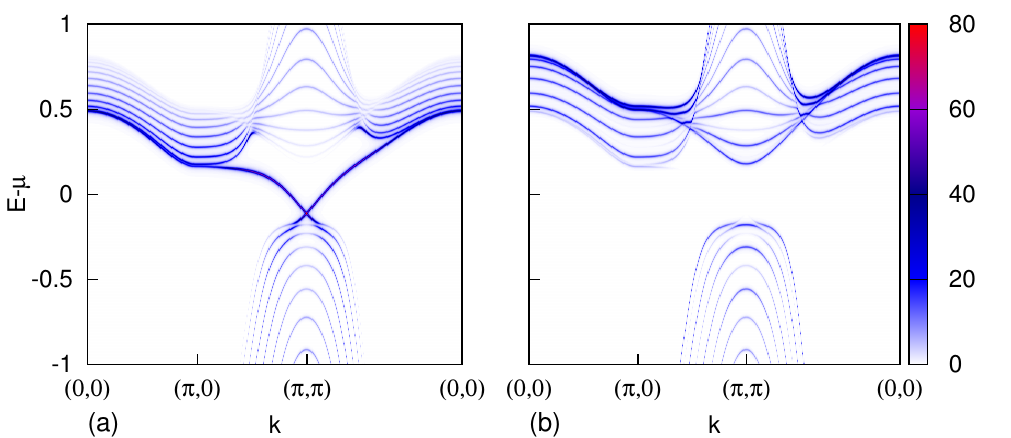}
\caption{Layer-resolved spectral intensity $A(\omega,\bk,z)$ \eqref{akslab} in the STI phase (for the same parameters as Fig. \ref{fig_bands_sti}): (a) on the surface ($z=1$) and, for comparison, (b) in the bulk ($z=8$).
}
\label{fig_arpes_sti}
\end{figure}

\subsubsection{Strong topological insulator}
\label{sec:cleanstrong}

For the STI phase we used parameters $t_c=1$, $t_f=-0.1$, $V=0.25$, $\epsilon_f=4$.
This choice (positive $\epsilon_f$, small $V$) stems from the necessity to have (i) a small Fermi velocity of the surface Dirac cone and (ii) a sizeable energy window around the Dirac point without additional surface states, in order to be able to study Dirac-cone QPI.
As a result, $T_K$ is much larger than the bandwidth due to strong valence fluctuations.
Despite this, the quasiparticle weight for surface states remains small, $Z_{\rm Dirac}\approx 0.07$.
Further we have $E_{\textrm{Dirac}}-\mu=-0.113$, $\dbulk=0.33$, bulk values of $b=0.90$, $\lambda=-0.44$, surface values $b_{z=1}=0.90$, $\lambda_{z=1}=-0.32$ and a surface Fermi velocity of $v=0.23$.
Band structure and layer-resolved intensities are shown in Figs. \ref{fig_bands_sti} and \ref{fig_arpes_sti}, with a single Dirac cone at $(\pi,\pi)$.



\subsubsection{Dirac cone spin structure}

For a full characterization of the surface states we have analyzed their spin--momentum locking, with details given in Appendix \ref{app_spin}. We find the STI Dirac cone to be described by the effective Hamiltonian
\begin{equation}
H_{STI}=v(k_y\sigma_x-k_x\sigma_y) \label{h_sti}
\end{equation}
where $(k_x,k_y)$ is measured from the center of the cone at $(\pi,\pi)$ -- this corresponds to the standard situation with spin perpendicular to momentum.\cite{tirev1}
The WTI Dirac cones have somewhat different spin structures, described by
\begin{equation}
H_{WTI}=\pm v(k_y\sigma_x+k_x\sigma_y) \label{h_wti}
\end{equation}
where momenta are measured from the centers of the cones at $(0,\pi)$ and $(\pi,0)$. This unusual spin-momentum locking should be measurable by spin-polarized photoemission experiments and is illustrated in Figs.~\ref{fig_qpi_wti}(a) and \ref{fig_qpi_sti}(a) below.


\section{Results: Dilute defects}
\label{sec:oneimp}


Kondo holes in Kondo insulators are known \cite{schlottmann_1, schlottmann_2} to create a bound state in the gap, or close to the band edge. We have verified that this also applies to Kondo holes in the bulk of a topological Kondo insulator, essentially because a bound state emerges generically\cite{kruis01,tkianderson} from strong scattering, provided that particle--hole symmetry is not too strongly broken, and is protected by the gap.
The situation becomes more interesting for a Kondo hole at or near the surface of a topological Kondo insulator, which is metallic, such that the bound state turns into a resonance which can be in principle observed by STM.

\begin{figure}[!tb]
\includegraphics[width=0.49\textwidth]{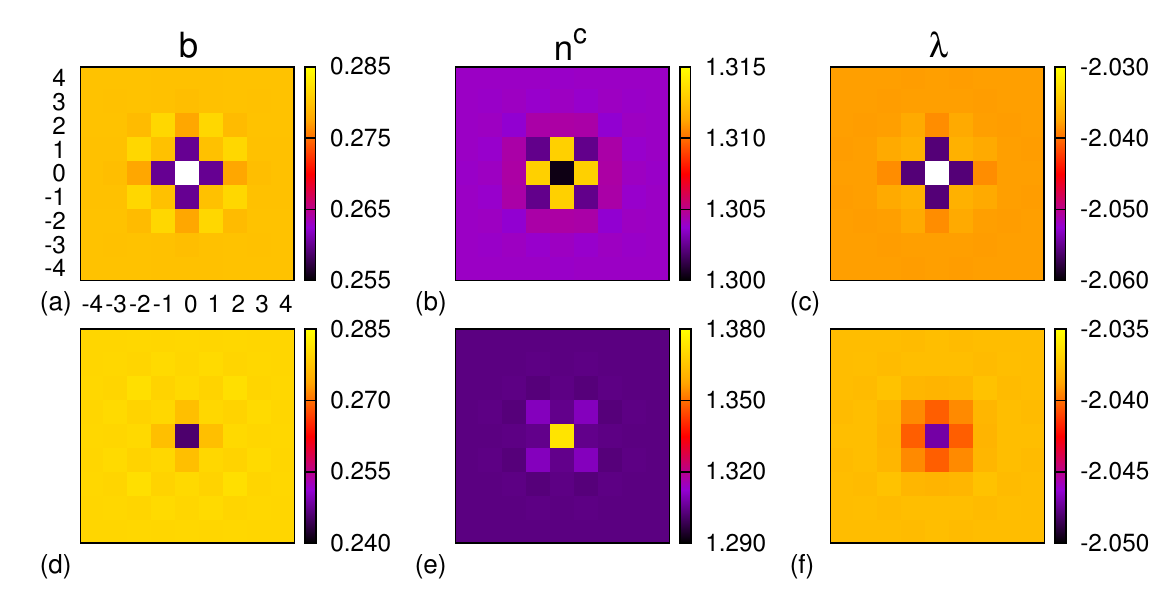}\\
\caption{Spatially resolved mean-field parameters near a Kondo hole, with model parameters in the WTI phase as in Fig.~\ref{fig_bands_wti}.
Shown are
(a,d): $b_i$,
(b,e): $n_i^c = \sum_\sigma \langle c_{i\sigma}^\dagger c_{i\sigma} \rangle$,
(c,f): $\lambda_i$,
in the surface ($z=1$) layer, for the two cases:
Upper panel: Kondo hole at site $(0,0)$ in the $z=1$ layer,
Lower panel: Kondo hole at site $(0,0)$ in the $z=2$ layer.
In (a-c) the $f$ orbital is missing at $(0,0)$, hence $b_i$ and $\lambda_i$ are not defined.
}\label{fig_scf}
\end{figure}

Therefore we now consider Kondo holes in either the surface layer or the layer below. To this end, we perform fully self-consistent mean-field calculations for a system of size $n_x\times n_x\times N_z$ in slab geometry, typically with $n_x=9$ and $N_z=15$. Sample results for the mean-field parameters in the case of a Kondo hole in the WTI phase are shown in Fig.~\ref{fig_scf}. These inhomogeneous mean-field parameters are then used as input for the T-matrix calculation, as described in Section~\ref{sec_embedding} above, to determine the LDOS and the QPI spectra from Kondo holes. These QPI spectra will be compared to those from weak impurities, as the latter allow us to gain some analytical insight.

\begin{figure}[!htb]
\includegraphics[width=0.49\textwidth]{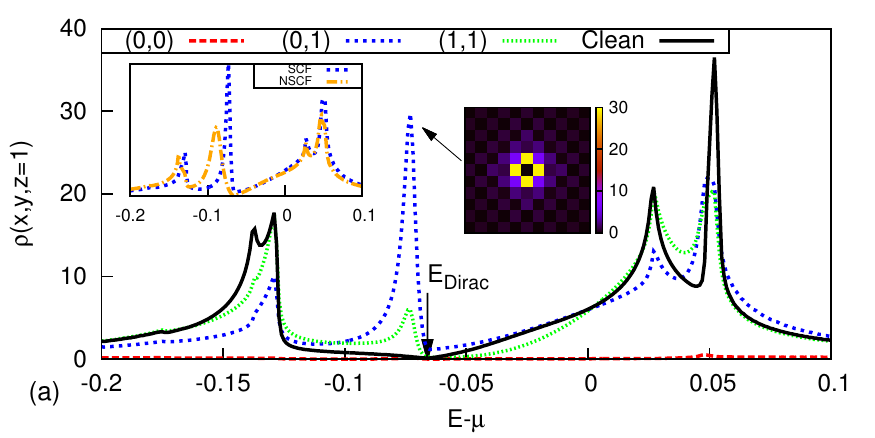}
\includegraphics[width=0.49\textwidth]{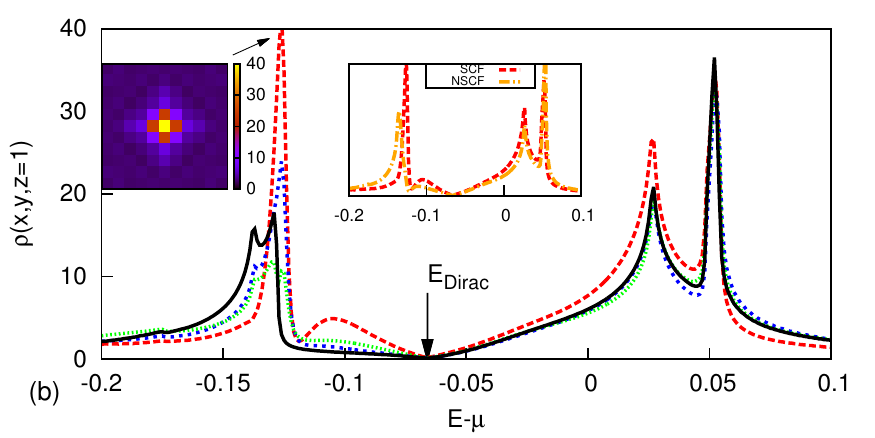}
\caption{LDOS $\rho(\w,x,y,z)$ from scattering matrix calculation (Eq. \eqref{eq_dos_t}) on the first layer $z=1$ in the WTI phase around a Kondo hole at (0,0) (a) in the first, or (b) in the second layer, for $N_x=801$.
Different curves show the LDOS for different sites $(x,y)$ close to the hole, the clean case is shown for comparison. The peaks around $-0.14$ and $0.05$ arise from van-Hove singularities of the surface states. The hole-induced resonance is below the Dirac energy, with its spatial intensity distribution illustrated in the color-scale insets.
The inset graphs show the difference in the resulting LDOS between a self-consistent (SCF) and non-self-consistent (NSCF) calculation, for the (0,1) site in (a) and for the (0,0) site in (b).
}\label{fig_dos_wti}
\end{figure}

\begin{figure}[!htb]
\includegraphics[width=0.49\textwidth]{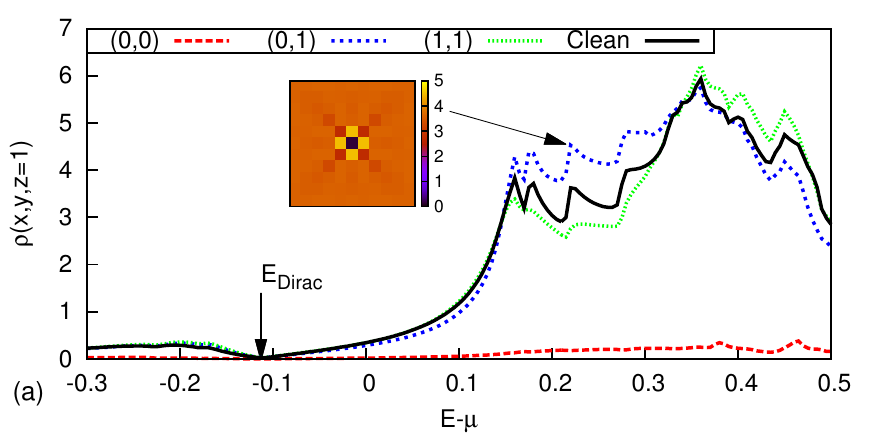}
\includegraphics[width=0.49\textwidth]{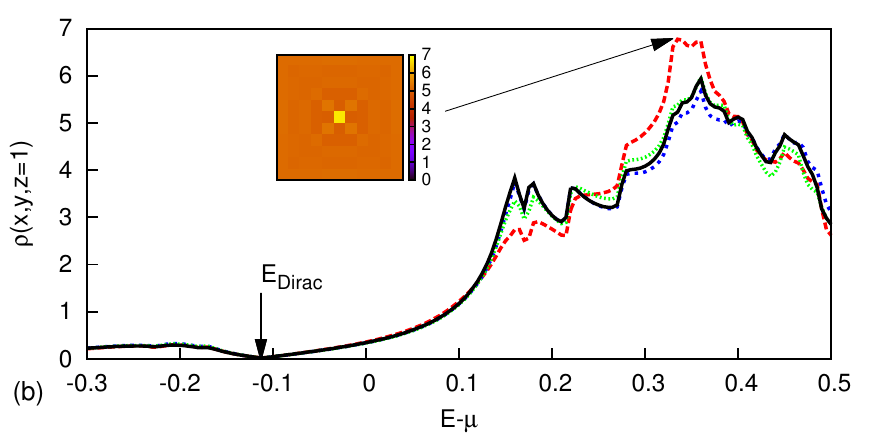}
\caption{Same as Fig.~\ref{fig_dos_wti}, but now for the STI phase.
In this case the bound state can be seen far from the Dirac point, hybridizing with bulk $f$-states, and does not appear as a clear resonance.
The difference between SCF and NSCF results is negligible here, since bulk states are very weakly affected by the hole.
}
\label{fig_dos_sti}
\end{figure}


\subsection{Local density of states (LDOS) for Kondo holes}

The surface-layer LDOS, detectable in an STM experiment and obtained from the T-matrix calculation, is shown in Fig. \ref{fig_dos_wti} for the weak topological Kondo insulator with parameters as in Section~\ref{sec:cleanweak}. For the case of a Kondo hole in the surface layer, Fig. \ref{fig_dos_wti}(a), we see that a resonance appears in the bulk gap, and it hybridizes with surface states. It is mainly localized on the four sites surrounding the hole, with a rapid spatial decay. Similar resonances have been predicted in non-Kondo TIs.\cite{balatsky_imp_ti, balatsky_imp_12,balatsky_imp_12_b}
For comparison, we have also performed a non-self-consistent calculation where the changes of mean-field parameters due to the impurity have been ignored, such that the T-matrix is non-zero on a single site only. The two results differ significantly concerning the energetic position of the resonance, underlining that full self-consistency is important.

For a hole in the second layer, Fig. \ref{fig_dos_wti}(b), the resonance appears weaker and at higher binding energy, close to the van Hove singularities of surface states. In the surface-layer LDOS, it is visible essentially only on the site above the hole. We note that the energetic location of the resonance depends on microscopic details, i.e., we have also encountered cases with a sharp low-energy resonance for a hole in the second layer.

In contrast, for the strong topological Kondo insulator with parameters as in Section~\ref{sec:cleanstrong} we have not found sharp low-energy resonances for any parameter set investigated. The reason is that the STI phase of model \eqref{H_r} only occurs in the mixed-valence regime, which in turn implies strong particle--hole asymmetry.
Since, for scattering in Dirac systems, the resonance energy $E_{res}$ is a function of both scattering strength and particle--hole asymmetry,\cite{kruis01,balatsky_imp_12_b}
determined essentially by $\R G^0(E_{res}) V=1$, increasing particle--hole asymmetry shifts the resonance of a strong scatterer away from the Dirac energy.
For our parameters, we only found minor impurity signatures in the low-energy LDOS, Fig.~\ref{fig_dos_sti}. Impurity-induced changes are visible at higher energies, but spoiled by the influence of bulk states.

\begin{figure}[!bt]
\includegraphics[width=0.49\textwidth]{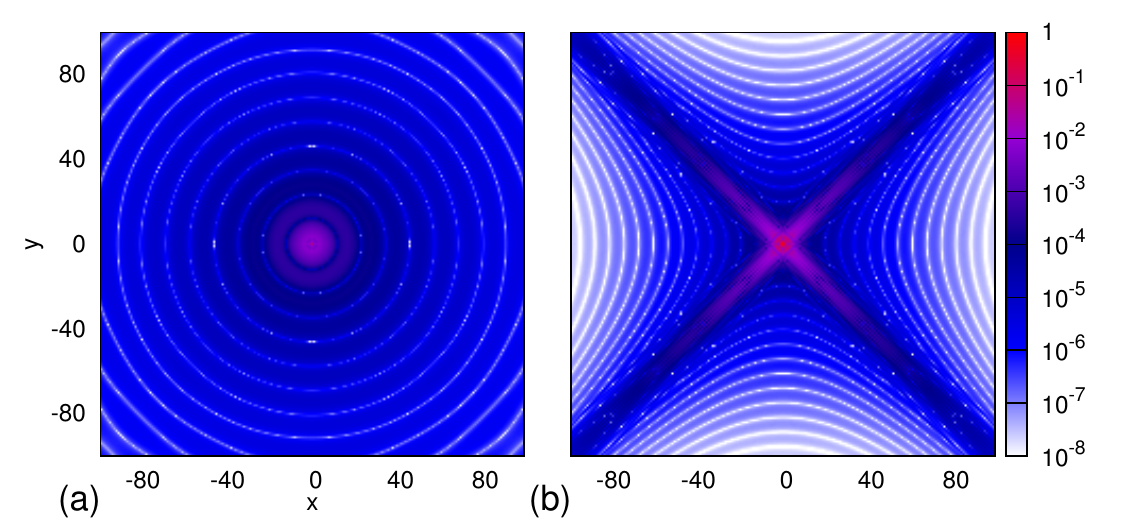}
\caption{Spatially resolved absolute value of the impurity-induced LDOS, $|\Delta \rho(\omega,x,y,z=1)|$ on a logarithmic scale around a Kondo hole in the first layer in the STI phase where $E_{\textrm{Dirac}}-\mu=-0.113$:
(a) at $E-\mu=-0.080$ for which isotropic Friedel oscillations can be observed,
(b) at $E-\mu=+0.120$ for which the nesting effect of the Fermi surface is strong, see Fig.\ref{fig_qpi_sti} (a4) below, giving rise to a ``focusing'' effect along the $(0,0)-(\pi,\pi)$ direction.
}\label{fig_stm}
\end{figure}

When analyzing the impurity-induced changes in the LDOS, $\Delta\rho$, as a function of the distance from the hole, Friedel oscillations with a wavelength $2\pi/(2\omega/v)=\pi v/\omega$ can be observed for energies close to $E_{\textrm{Dirac}}$, Fig. \ref{fig_stm}(a).
As long as warping effects can be neglected, the decay is isotropic, and proportional to $r^{-1}$ in the WTI phase and to $r^{-2}$ in the STI phase,
in agreement with earlier results for graphene \cite{bena_graphene, mariani_graphene} and for STIs.\cite{kim_friedel, wang_ti_imp,liu_qpi_spa}
At higher energies, when warping effects cannot be neglected, the decay becomes anisotropic, and a strong focusing effect can be observed if a nesting of the Fermi surface can be achieved, Fig. \ref{fig_stm}(b).

\subsection{QPI from weak impurities}

Before analyzing the quasiparticle interference signal caused by a Kondo hole, it is useful to analyze a few QPI properties of weak impurities, where we neglect any impurity-induced changes of mean-field parameters. A comparison of numerical QPI results for both Kondo holes and weak impurities can be found in Figs. \ref{fig_qpi_wti} and \ref{fig_qpi_sti} below.

In the following discussion we restrict our attention to energies within the bulk gap. This enables an analytical treatment close to the Dirac points, using the effective surface Hamiltonian Eqs. \eqref{h_sti} or \eqref{h_wti}.
%
This approach has been taken in the literature before, and we start with reviewing these results. In what follows we measure energies relative to the Dirac energy: $\omega\equiv E- E_{\textrm{Dirac}}$.

\subsubsection{Point-like effective impurity}

We focus first on a strong topological insulator with the surface Hamiltonian \eqref{h_sti}.
The unperturbed surface Green's function is 
\begin{equation}
\hat G^0(\omega,\bk)=\frac{1}{\omega^2-v^2k^2}(\omega \hat\sigma_0 +vk_y\hat\sigma_x -vk_x\hat\sigma_y).
\end{equation}
For scatterers which act as point-like impurities in the effective theory -- this assumption is in general not justified, see below -- the T-matrix is momentum-independent, and the scattering matrix equation \eqref{gg0t} gives, for intracone scattering,
\begin{equation}
\Delta \rho_{intra}(\omega,\bq)=-\frac{1}{\pi}\I\Tr\sum_{\bk}\hat G^0(\omega,\bk)\hat T(\omega)\hat G^0(\omega,\bk-\bq).
\end{equation}
Here we have used that $\Delta \rho(\omega,\bq)$ is real for the single-impurity case considered here. For a non-magnetic impurity, both $\hat V$ and $\hat T(\omega)$ are proportional to the identity in the spin space, so can be treated as scalar quantities, and we get:
\begin{equation}
\Delta \rho_{intra}(\omega,\bq)=-\frac{1}{\pi}\I\left[T_{intra}(\omega)\Lambda_{intra} (\omega,\bq)\right]\label{qpi_intra1}
\end{equation}
where
\begin{equation}
\Lambda_{intra}(\omega,\bq)=\int \frac{d^2\bk}{(2\pi v)^2}\frac{\omega^2+v^2k^2-v^2\bk\cdot\bq}{(\omega^2-v^2k^2)[\omega^2-v^2(\bk-\bq)^2]},\label{qpi_intra2}
\end{equation}
and we have made explicit that $T(\omega)$ describes intracone scattering $T_{intra}(\omega)$.
This result has been derived before in the context of graphene,\cite{pereg_graphene} STIs,\cite{guo_franz_sti} high-temperature superconductors,\cite{franz_superc}
by going into Matsubara frequencies $\omega \rightarrow i\omega$, applying the Schwinger-Feynman parametrization trick, then coming back to real frequencies.
It only depends on the magnitude $q=|\bq|$ of the transferred momentum $\bq$, and, up to a momentum-independent additive term that we omit, reads:
\begin{equation}
\Lambda_{intra}(\omega,\bq)=-\frac{1}{\pi v^2} F\left(\frac{qv}{2\omega}\right), \label{lambda_intra_f}
\end{equation}
with
\begin{equation}
F(z)=\frac{\sqrt{1-z^2}}{z}\arctan \frac{z}{\sqrt{1-z^2}}+i\frac{\pi}{2}. \label{f(z)}
\end{equation}

\begin{figure}[!tb]
\includegraphics[width=0.48\textwidth]{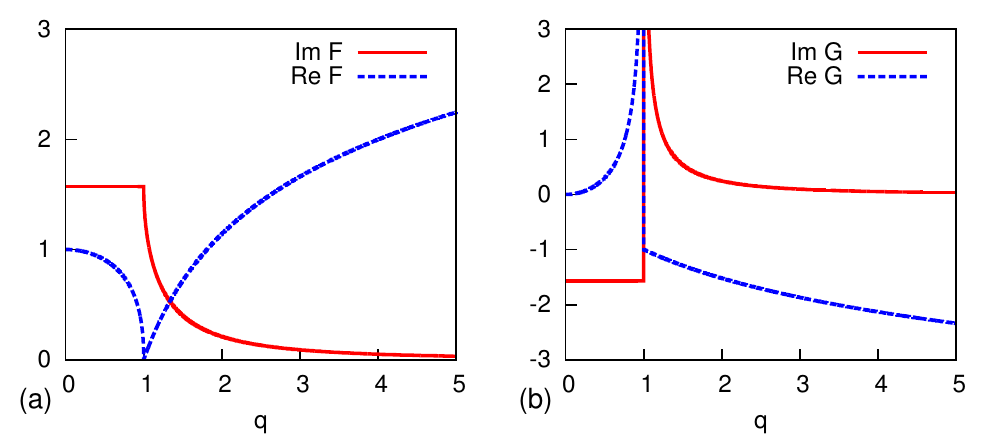}\\
\caption{The real and imaginary parts of functions (a) $F(q)$, Eq. \eqref{f(z)}, and (b) $G(q)$, Eq. \eqref{g(z)}, which describe, respectively, intra- and intercone scattering in the Born approximation.
}\label{fig_fg}
\end{figure}

Now we turn to a weak topological insulator, with two Dirac cones described by the effective Hamiltonian Eq. \eqref{h_wti}.
The surface Green's function becomes
\begin{equation}
\hat G^0_{\pm}(\omega,\bk)=\frac{1}{\omega^2-v^2k^2}[\omega \hat\sigma_0 \pm v(k_y\hat\sigma_x +k_x\hat\sigma_y)]\label{g+-_wti}
\end{equation}
The intracone scattering leads to the same expressions Eqs. \eqref{qpi_intra1}, \eqref{qpi_intra2} for both cones.
The intercone scattering, instead, leads to:
\begin{align}
&\Delta \rho_{inter}(\omega,\bq)=\nonumber\\
&=-\frac{1}{\pi}\I\Tr\sum_{\bk}\hat G^0_+(\omega,\bk)\hat T(\omega)\hat G^0_-(\omega,\bk-\bq-\bQ) =\nonumber\\
&=-\frac{1}{\pi}\I\left[T_{inter}(\omega)\Lambda_{inter} (\omega,\bq+\bQ)\right]\label{qpi_inter1}
\end{align}
where $\bQ=(\pi,\pi)$ is the distance between the cones, and
\begin{equation}
\Lambda_{inter}(\omega,\bq)=\int \frac{d^2\bk}{(2\pi v)^2}\frac{\omega^2-v^2k^2+v^2\bk\cdot\bq}{(\omega^2-v^2k^2)[\omega^2-v^2(\bk-\bq)^2]}.\label{qpi_inter2}
\end{equation}
Interestingly, this function also appears in the description of intracone scattering by a {\em magnetic} impurity in a STI\cite{guo_franz_sti} or in a high-temperature superconductor,\cite{franz_superc}
or of intracone scattering in graphene by a staggered potential,\cite{pereg_graphene} or of intercone scattering in graphene, here up to a direction-dependent factor.\cite{pereg_graphene} It is:
\begin{equation}
\Lambda_{inter}(\omega,\bq)=-\frac{1}{\pi v^2} G\left(\frac{qv}{2\omega}\right),
\end{equation}
with
\begin{equation}
G(z)=\frac{z}{\sqrt{1-z^2}}\arctan \frac{z}{\sqrt{1-z^2}}-i\frac{\pi}{2}. \label{g(z)}
\end{equation}
The functions $F(z)$ and $G(z)$ are shown in Fig. \ref{fig_fg}.

In the Born approximation, $T(\omega)=V$ is real, so the signal is proportional to the imaginary part of $\Lambda$ functions, and, for a WTI, is given by the sum of three contributions (two intracone, and one intercone):
\begin{equation}
\Delta \rho_{WTI} (\omega,\bq)= 2 \Delta \rho_{intra}(\omega, \bq) + \Delta \rho_{inter}(\omega, \bq),
\end{equation}
while for a STI it is simply the intracone signal:
\begin{equation}
\Delta \rho_{STI} (\omega,\bq)= \Delta \rho_{intra}(\omega, \bq),
\end{equation}
with
\begin{eqnarray}
\label{vintra}
\Delta \rho_{intra}(\omega,\bq)&=&-\frac{1}{\pi}V_{intra}\I\left[\Lambda_{intra} (\omega,\bq)\right],\\
\label{vinter}
\Delta \rho_{inter}(\omega,\bq)&=&-\frac{1}{\pi}V_{inter}\I\left[\Lambda_{inter} (\omega,\bq+\bQ)\right],
\end{eqnarray}
where we have taken into account that a given microscopic scatterer will give rise to different intra- and intercone scattering amplitudes $V_{inter}$ and $V_{intra}$.
It can be observed that the imaginary part of $\Lambda_{intra}$ is flat for $vq/2\omega<1$, with a kink and no divergence at $vq/2\omega=1$.
This is a well-known consequence of the inhibited backscattering by non-magnetic impurities in a TI cone,\cite{tirev1} due to the opposite direction of the spin when $\bk\leftrightarrow (-\bk)$, see Fig. \ref{fig_qpi_sti}(a2).
In contrast, the imaginary part of $\Lambda_{inter}$ diverges for $vq/2\omega=1^+$: in this case, intercone scattering by this wavevector leads to a final state with the same spin as the initial state, see Fig. \ref{fig_qpi_wti}(a2).

\subsubsection{Extended effective impurity}

For extended scatterers where the scattering matrix element in the effective theory depends on the transferred momentum $\bq$ only, the Born-limit analytical results are modified into\cite{capriotti,guo_franz_sti}
\begin{equation}
\Delta \rho(\omega,\bq)=-\frac{1}{\pi}|V_\bq| \I \Lambda(\omega,\bq)
\label{vqborn}
\end{equation}
where $|V_\bq|$ is the Fourier-transformed scattering profile which now modulates the QPI signal.

\subsubsection{Point-like microscopic impurity}
\label{sec:pointmic}

To bridge between microscopic and effective modelling, microscopic scattering terms need to be transformed into those for the surface theory. First, this implies that impurities which are microscopically located in the $c$ and $f$ bands cause QPI signals of different amplitude, because the surface states have largely $f$ character and hence couple more strongly to $f$ impurities.

Second, the transformation between microscopic and effective degrees of freedom is general momentum-dependent. This implies that an impurity which is point-like within the microscopic model, i.e., acts on a single lattice site, is {\em not} point-like in the effective theory -- a problem which is sometimes overlooked in the literature.
In other words, a microscopic scattering term $\sum_{\bk,\bk'} c_\bk^\dagger c_{\bk'}$ transforms into $\sum_{\bk,\bk'} u^\ast_\bk u_{\bk'} a_\bk^\dagger a_{\bk'}$ in terms of effective particles $a_\bk$.
The resulting scattering matrix element does not only depend on the transferred momentum $(\bk-\bk')$. Hence, Eqs.~(\ref{vintra},\ref{vinter}) and also Eq.~\eqref{vqborn} are not applicable beyond the limit of small $\bq$, i.e., the momentum dependencies of scatterer and band structure mix.

\begin{figure*}[!tb]
\includegraphics[width=0.8\textwidth]{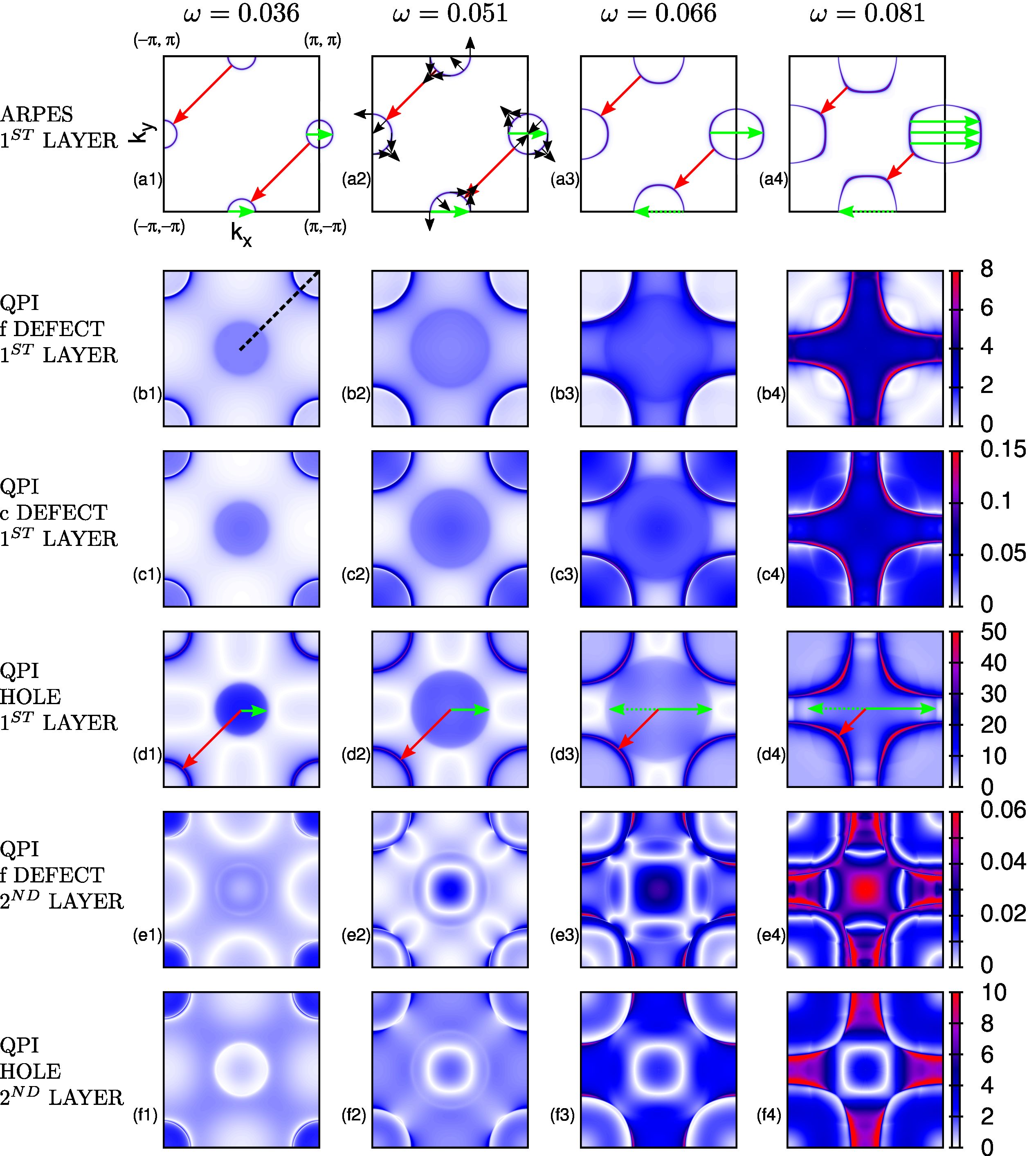}\\
\caption{
Constant-energy cuts through
(a) the single-particle spectrum $A(\omega,\bk,z=1)$ and
(b-f) the surface QPI signal $|\rqpi(\w,\bq)|$
in the WTI phase with parameters as in Fig.~\ref{fig_bands_wti}.
The columns correspond to energies $\omega=0.036$, $0.051$, $0.066$ and $0.081$ relative to the Dirac point $E_{\textrm{Dirac}}-\mu=-0.066$.
Panel (a) also shows the spin directions of the surface-state electrons (black arrows) and the relevant intracone (green) and intercone (red) scattering wavevectors.
The QPI panels display the response of
(b) a weak $f$-scatterer ($\Delta \epsilon_f=0.01$) in the surface layer ($z=1$), 
(c) a weak $c$-scatterer ($\Delta \epsilon_c=0.01$) at $z=1$, 
(d) a Kondo hole at $z=1$,
(e) a weak $f$-scatterer in the second layer ($z=2$), and 
(f) a Kondo hole at $z=2$.
The dashed line in (b1) shows the momentum-space cut for which $\rqpi(\w,\bq)$ is shown in Fig.~\ref{fig_qpi_wti_1d}.
}\label{fig_qpi_wti}
\end{figure*}

\begin{figure*}[!tb]
\includegraphics[width=0.8\textwidth]{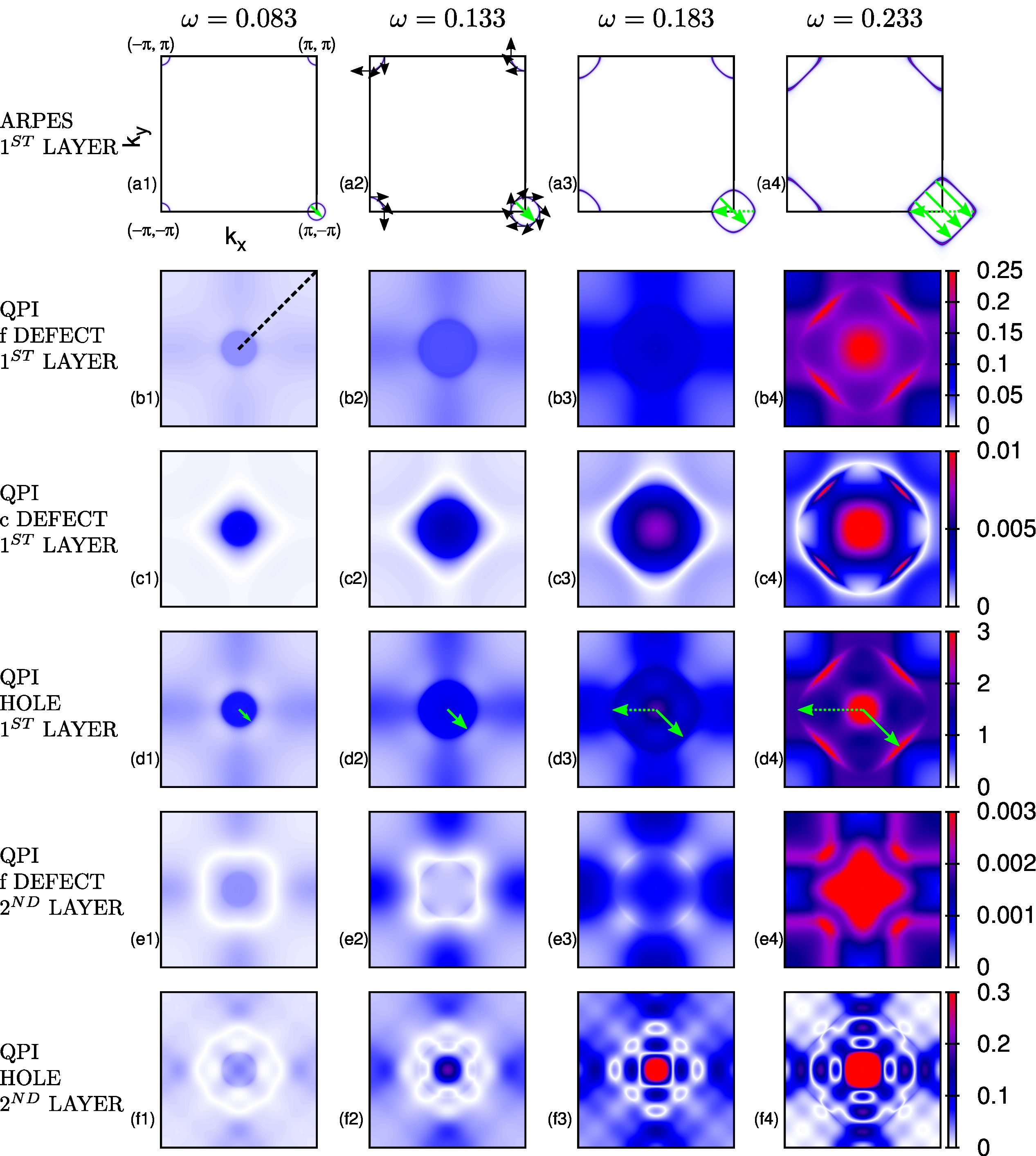}\\

\caption{Same as Fig.~\ref{fig_qpi_wti}, but now in the STI phase with parameters as in Fig.~\ref{fig_bands_sti} and $E_{\textrm{Dirac}}-\mu=-0.113$.
As intercone scattering processes are absent, the QPI signal is weak due to forbidden backscattering
unless warping/nesting effects play a role.
}\label{fig_qpi_sti}
\end{figure*}

This is nicely seen from the QPI results for weak point-like impurities in the $c$ and $f$ band, obtained from a full microscopic calculation. Comparing Figs.~\ref{fig_qpi_wti}(b) and (c) one notices that the QPI patterns for both cases, here in the WTI phase, show a different momentum dependence, particularly pronounced away from the Dirac point. The same applies to Figs.~\ref{fig_qpi_sti}(b) and (c), which illustrate this point for the STI phase.

\begin{figure*}[!tb]
\includegraphics[width=0.9\textwidth]{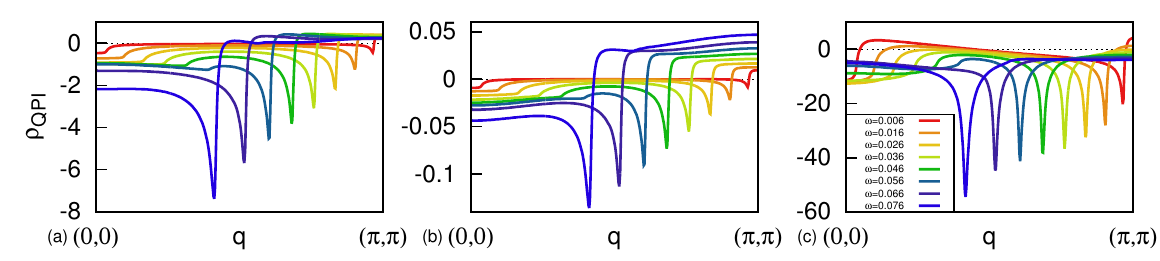}\\
\caption{Momentum-space cuts along $(0,0)-(\pi,\pi)$ through the QPI signal $\rqpi(\w,\bq)$
at different energies in the WTI phase with parameters as in Fig.~\ref{fig_bands_wti}, for
(a) a weak $f$-scatterer ($\Delta \epsilon_f=0.01$) in the surface layer ($z=1$), 
(b) a weak $c$-scatterer ($\Delta \epsilon_c=0.01$) at $z=1$, 
(c) a Kondo hole at $z=1$.
For cases (a) and (b) the signal close to $\bq=0$ resembles $\I F(qv/2\omega)$, Fig.~\ref{fig_fg}(a), while the signal close to $\bq=(\pi,\pi)$ follows $\I G(qv/2\omega)$, Fig. \ref{fig_fg}(b).
}\label{fig_qpi_wti_1d}
\end{figure*}

\begin{figure*}[!tb]
\includegraphics[width=0.9\textwidth]{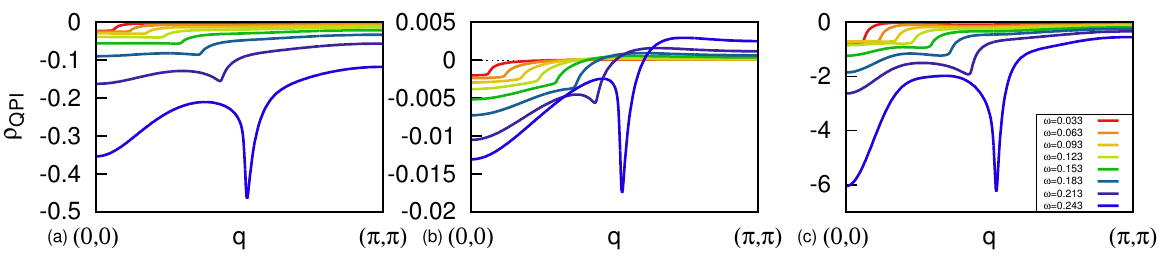}\\
\caption{Same as Fig. \ref{fig_qpi_wti_1d}, but now in the STI phase with parameters as in Fig.~\ref{fig_bands_sti}.
}\label{fig_qpi_sti_1d}
\end{figure*}

\subsubsection{Beyond the Born approximation}

Beyond the Born approximation, $T(\omega)$ can be computed through
\begin{equation}
T(\omega)=\frac{V}{1-G^0(\omega) V},
\end{equation}
with
\begin{equation}
G^0(\omega)=\frac{1}{2}\int \frac{d^2\bk}{(2\pi)^2} \Tr\hat G^0(\omega,\bk),
\end{equation}
and is a complex quantity, which can be written as
\begin{equation}
T(\omega)=|T(\omega)|e^{i\delta(\omega)}
\end{equation}
by introducing the phase shift $\delta(\omega)$, which is zero in the Born approximation. As a consequence, Eqs.~(\ref{qpi_intra1},\ref{qpi_inter1}) can be written as:
\begin{align}
&\Delta \rho (\omega,\bq)=-\frac{1}{\pi} |T(\omega)| \cdot\nonumber\\
&\cdot[\cos\delta (\omega) \I\Lambda(\omega,\bq)+\sin\delta (\omega) \R\Lambda(\omega,\bq)],
\end{align}
which shows that a linear combination of the imaginary and real parts of the $\Lambda$ functions is observed, depending on the energy-dependent phase shift $\delta(\omega)$.

Since $\R F(z)$ has a cusp for $z=1$, the intercone signal may now show a shallow maximum (or minimum) at $q=2\omega/v$; moreover, while $\I G(z)$ diverges for $z\rightarrow 1^+$, $\R G(z)$ diverges for $z\rightarrow 1^-$, so the intercone signal always diverges for $|\bQ-\bq|=2\omega/v$, and can in principle show any kind of interference-like pattern.

\subsubsection{Beyond the Dirac-cone approximation: Warping}

For energies sufficiently far from the Dirac point the isotropic effective theory of Eqs.~\eqref{h_sti} and \eqref{h_wti} is no more valid, and a square warping effect can be observed, similar to the hexagonal warping of Bi$_2$Se$_3$ and related compounds, see Refs. \onlinecite{bi2te3}, \onlinecite{fu_warping}, and \onlinecite{lee_qpi}.

In the STI phase warping causes a strong peak along the $(0,0)-(\pi,\pi)$ direction even in the Born approximation, since it allows for nesting of the Fermi surface, see Fig. \ref{fig_qpi_sti}(b4) and (c4).
Warping happens in the WTI phase, too, even though a strong nesting effect cannot be observed.
In general the exact details of the warping effect depend on the choice of the parameters, so it is not possible to make universal predictions in this regime.


\subsection{QPI from Kondo holes}

We now turn to the discussion of the QPI results for Kondo holes, where $V$ and $T(\omega)$ extend over all the sites of the $n_x\times n_x\times n_z$ (practically $9\times9\times3$) scattering region. This calculation needs to be done numerically as described in Section~\ref{sec_embedding}.

QPI results for an isolated Kondo hole in the surface layer are shown in Figs.~\ref{fig_qpi_wti}(d) and \ref{fig_qpi_sti}(d) for the WTI and STI phases, respectively. Comparing the Kondo-hole case to the weak-impurity cases, Figs.~\ref{fig_qpi_wti}(b,c) and \ref{fig_qpi_sti}(b,c), one notices that
(i) peak shapes change as the phase of the $T$ matrix becomes relevant and
(ii) weight in momentum space is redistributed due to the stronger momentum dependence of the scattering potential.

\begin{figure}[!b]
\includegraphics[width=0.48\textwidth]{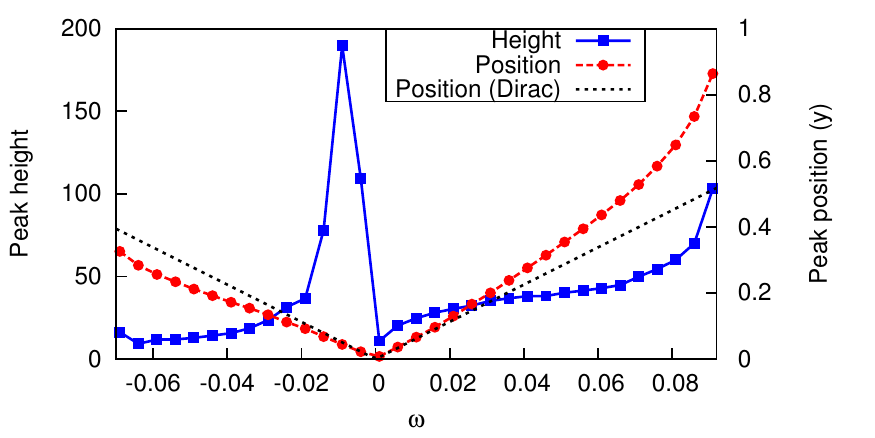}\\
\caption{Energy-dependent intensity and $\bq$-space position of the intercone QPI peak for a $z=1$
Kondo hole in the WTI phase, corresponding to Fig.~\ref{fig_qpi_wti_1d}(c).
The intensity is roughly proportional to the vacancy-induced LDOS, see Fig. \ref{fig_dos_wti}(a), while it would simply follow the bulk LDOS in the Born approximation (not shown).
The peak position $y$ is parameterized as $\bq=(1-y)(\pi,\pi)$ such that $y=0$ corresponds to intercone scattering at the Dirac energy, $\w=0$. In the Dirac-cone approximation we have
$y=(2|\omega|/v)/(\pi\sqrt{2})$, with the Fermi velocity $v=0.08$ (black dotted).
}\label{fig_qpi_1d_peak}
\end{figure}

Both effects are more clearly seen when looking at momentum-space cuts along $(0,0)-(\pi,\pi)$ through the QPI spectrum, shown in Figs.~\ref{fig_qpi_wti_1d} and Fig. \ref{fig_qpi_sti_1d}. For example, the shape of the intercone peak in Figs.~\ref{fig_qpi_wti_1d}(a) and (c) is rather different. For the Kondo hole, one moreover observes a strong energy dependence (resonant in the WTI case) of the QPI signal which roughly follows the LDOS, as can be seen by comparing Figs.~\ref{fig_qpi_1d_peak} and Fig.~\ref{fig_dos_wti}(a).

Upon comparing the WTI and STI phases, the results show that the QPI signal for intracone scattering (green arrows) is weak and not peaked: This is simply a result of backscattering being suppressed by spin-momentum locking. (Recall that this is roughly described by $\I F$ shown in Fig.~\ref{fig_fg}.)
In contrast, for intercone scattering there are wavevectors (red arrows) connecting states with equal spin, resulting in a larger and strongly peaked QPI signal.
As function of energy the crossover from nearly isotropic Fermi surface and QPI signal (left column) to a signal dominated by square warping (right column) is apparent, with approximate nesting leading to a large gain in QPI intensity.

We have also calculated QPI patterns for impurities not located in the surface ($z=1$) layer, but beneath it. In general, the low-energy QPI signal is weaker here as compared to surface impurities, simply because the surface states live mainly in the $z=1$ layer. In panels (e) and (f) of Figs.~\ref{fig_qpi_wti} and \ref{fig_qpi_sti} we show corresponding results for a weak impurity in the $f$ band and a Kondo hole, respectively, both located in the $z=2$ layer. While the main qualitative features remain, the QPI patterns tend to be more complicated, as now the full spatial matrix structure of the $T$ matrix becomes important.

Taking the panels in Fig.~\ref{fig_qpi_wti} together, we see that the QPI signal displays a large variation between the different types of impurities -- the same applies to Fig.~\ref{fig_qpi_sti}. Even weak point-like impurities in different bands produce different QPI patterns, related to their non-local character in the effective surface theory, see Section~\ref{sec:pointmic}. On the one hand this certainly complicates the interpretation of QPI experiments, but on the other hand may be exploited to characterize experimentally existing defects on the basis of careful modelling.

\subsection{QPI from magnetic impurities: Qualitative discussion}

All results so far concern the non-magnetic (spin-singlet) QPI response to non-magnetic (spin-singlet) impurities. As shown in Ref.~\onlinecite{madhavan_magimp_ti}, magnetic impurities allow QPI to probe scattering channels which are otherwise prohibited by spin-momentum locking when time reversal symmetry is preserved.

In this subsection we therefore quickly discuss both the magnetic (i.e. spin-antisymmetric or spin-triplet) response as well as magnetic impurities in the Dirac-cone approximation.\cite{guo_franz_sti} First, the spin-triplet QPI signal from non-magnetic impurities is always zero due to time reversal.

Magnetic impurities, instead, can lead to both triplet and singlet response. As shown in Ref. \onlinecite{guo_franz_sti}, the latter is zero in the Born approximation, while it is finite beyond. For a single Dirac cone, it is described by the $F(z)$ function, Eq.~\eqref{f(z)}, so it is small and non-diverging. The former, however, is always non-zero, and it is described by the $G(z)$ function, Eq.~\eqref{g(z)}, up to a direction-dependent factor, so it is strong and diverging at $|\bq|=2\omega/v$. These conclusions naturally apply to the STI phase of the topological Kondo insulator.

What happens for intercone scattering, relevant to the WTI phase?
For simplicity we consider a magnetic impurity polarized along the $z$ direction and calculate the triplet component of the LDOS polarized in $z$ direction:
\begin{align}
&\Delta\rho_{inter}^{\sigma_z\sigma_z}(\omega, \bq+\bQ)=\nonumber\\
&=-\frac{1}{\pi}\I \left[ T(\omega) \sum_\bk \Tr [\hat\sigma_z \hat G^0_+(\omega, \bk)\hat\sigma_z \hat G^0_-(\omega, \bk-\bq)]\right].
\end{align}
With $\hat G^0_{\pm}(\omega,\bk)$ from Eq.~\ref{g+-_wti} we see that:
\begin{equation}
\Delta\rho_{inter}^{\sigma_z\sigma_z}(\omega, \bq)=-\frac{1}{\pi}\I[T(\omega)\Lambda_{intra} (\omega, \bq+\bQ)],
\end{equation}
so the signal is proportional to $\Lambda_{intra} (\omega, \bq)$, Eq.~\eqref{qpi_intra2}, thus to function $F(z)$, Eq.~\eqref{lambda_intra_f}.
Hence, it has the same momentum structure as the intracone (!) scattering in the non-magnetic case, the only difference being that it is centered at $\bQ=(\pi,\pi)$ rather that at $(0,0)$.

When considering different magnetization and/or probe axis, the signal will be modulated according to the angle in the $xy$ plane as shown in Ref. \onlinecite{guo_franz_sti}, but still described by the function $F(z)$, i.e., is weak and non-diverging.

To summarize, for the magnetic response from a magnetic impurity the functions $F(z)$ and $G(z)$ switch their role w.r.t. the spin-unpolarized case:
$F(z)$ describes intracone singlet--singlet response and intercone triplet--triplet response, while
$G(z)$ describes intercone singlet--singlet response and intracone triplet--triplet response
(all in the isotropic Dirac cone approximation).
The intracone scattering signal is always centered at $(0,0)$, the intercone one at $(\pi,\pi)$.


\section{Results: Finite concentration of impurities}
\label{sec:manyimp}

In this section, we depart from the dilute limit of isolated impurities and turn to discuss the physics of a finite defect concentration. A treatment thereof requires no modifications to the real-space mean-field approach of Section~\ref{sec:rmf}, but energy and momentum resolution of the numerical results are now restricted by the system-size limit for the (self-consistent) diagonalization of the mean-field Hamiltonian \eqref{F_r} (amplified by the use of supercells).

\begin{figure}[!tbp]

\includegraphics[width=0.41\textwidth]{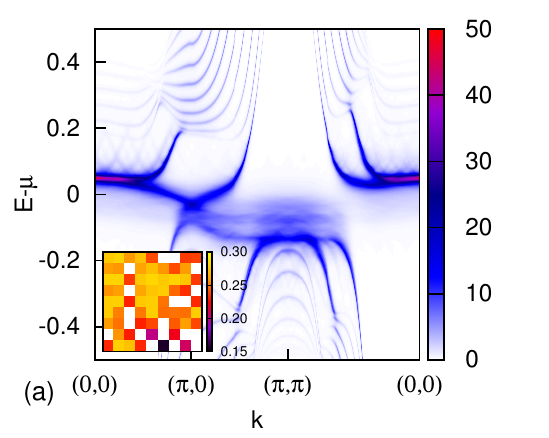}
\includegraphics[width=0.41\textwidth]{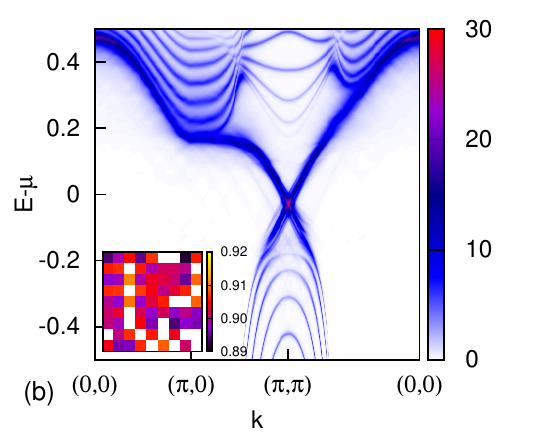}
\caption{Spectral intensity $A(\omega,\bk,z=1)$ of a $n_x\times n_x \times N_z$ supercell with a random surface concentration $\nimp=0.2$ of holes
for the same path as in Fig. \ref{fig_bands_wti},
(a) in the WTI phase, to be compared with Fig. \ref{fig_arpes_wti}, which is the case with no holes,
(b) in the STI phase, to be compared with Fig. \ref{fig_arpes_sti}.
We employed $n_x=9$, $N_z=15$, and an artificial broadening $\delta=0.005$.
The spectra have been averaged over 20 realizations of disorder.
In the inset we show the spatial map of $b_i$ on the surface for a particular realization of disorder (the same for both WTI and STI): white spots are the hole sites.
}
\label{fig_arpes_dis}
\end{figure}

We have performed simulations for a finite concentration of Kondo holes randomly distributed over all sites of the system as well as for a finite concentration of surface Kondo holes. In both cases and in the WTI phase, there is a clear tendency of the resonances around each impurity to form a low-energy band. This is illustrated in Fig. \ref{fig_arpes_dis} which displays the momentum-resolved single-particle spectrum of the surface layer for a concentration of $\nimp=\Nimp/N_x^2=0.2$ surface holes, where the impurity band around $E-\mu=-0.08$, superposed to the Dirac cones, is clearly visible. In addition, the overall low-energy intensity decreases roughly according to $(1-\nimp)$.

In the STI phase, instead, this effect is largely absent, since no appreciable change of the LDOS around the Dirac energy is caused by the holes.
At low energies, the only visible effect is a smearing of the Dirac cone, due to the local shift of the Dirac energy caused by disorder. At elevated energies impurity-induced weight can be found, but in $A(\omega,\bk,z=1)$ this weight is difficult to separate from that of bulk states due to the absence of $k_z$ resolution.

Once again, we note that this difference between STI and WTI phases is not dictated by topology, but by the different degree of particle--hole asymmetry in the two cases.
More results for a finite concentration of Kondo holes, together with a detailed analysis, will be presented in subsequent work.

\section{Conclusions}
\label{sec:concl}

This paper was devoted to a mean-field-based study of impurities in topological Kondo insulators, both in the WTI and STI phases. Our primary focus was on Kondo holes, i.e., sites with missing $f$-orbital degree of freedom, but we have also considered weak scatterers in both $c$ and $f$ channels. Most calculations were performed in a slab geometry, in order to study the effects of impurities on the observable properties of surface states, measurable via photoemisson or scanning-tunnelling spectroscopy. All results have been obtained in the low-temperature limit, $T\ll T_K,\dbulk$, and more expensive numerical techniques are required to study the full temperature dependence.\cite{assaad_temperature,benlagra11}

We find that in the WTI phase a Kondo hole creates a resonance, localized on the sites near the hole, close to the Dirac energy.
In the STI phase, due to the large particle--hole asymmetry, this resonance moves to higher energies, mainly hybridizing with bulk states.

QPI patterns are very different in the two phases, due to the different number of Dirac cones which are involved.
In the STI phase only a single Dirac cone is present, and, as widely shown in the literature, backscattering is suppressed due to spin-momentum locking, so the QPI signal is weak.
This is to be contrasted with the WTI phase, where intracone scattering remains suppressed but intercone scattering is not, so the resulting QPI signal is strong.
This results can be nicely justified analytically in the Born approximation for isotropic cones.
In the general case, full numerical calculations show  that different scatterers give rise to specific QPI patterns which in principle can be used to distinguish experimentally different kinds of impurities. In this context, we have pointed out that point-like microscopic impurities do, in general, not correspond to point-like impurities in effective surface theories.

We have also examined a finite concentration of Kondo holes and determined momentum resolved spectral functions. In the WTI phase the resonances around each hole form a broad impurity band near the Dirac energy, partly destroying the Dirac cone structure, while in the STI phase minor disorder broadening of the Dirac cone, combined with impurity-induced weight at elevated energies, is observed. This difference can be attributed not to topology, but to the different amount of particle--hole asymmetry.
Further work is required to study possibly emergent non-Fermi-liquid behavior\cite{kaul07} due to disorder.

Our predictions can be verified in future experiments; we also note that some results, e.g. on intercone QPI, are not specific to TKIs and thus of relevance beyond Kondo systems.


{\it Note added:} Upon completion of this manuscript, a related paper \cite{wangzhu14} appeared, which discusses isolated impurities in topological Kondo insulators in the framework of the Gutzwiller approximation. The results presented in Ref.~\onlinecite{wangzhu14} are compatible with ours, but do neither cover surface QPI nor finite impurity concentration.


\acknowledgments

We thank H. Fehske, L. Fritz, and D. K. Morr for discussions and collaborations on related work.
This research was supported by the DFG through FOR 960 and GRK 1621
as well as by the Helmholtz association through VI-521.


\appendix

\section{Equivalence of free-particle models}
\label{app_sitte}

The TKI model that we have used in this paper, originally proposed in Refs. \onlinecite{tki1,tki2}, reduces to a free-particle model at the mean-field level. This model turns out to be equivalent to the common cubic-lattice four-band model\cite{qi2008,sitte_ti} for 3D topological insulators.

This can be shown as follows:
If we set $t_c=-t_f\equiv t$, $b=1$, $\epsilon_f=-\epsilon_c\equiv -\epsilon$, $\lambda=\mu=0$, and rescale the hybridization along $z$ by a factor 2 through $V_z=-V_x/2=-V_y/2\equiv -\lambda/2<0$
(in this way the model acquires cubic symmetry),
Eq. \eqref{F_k} can be written as $H=\sum_\bk \Psi_\bk^\dagger H_k \Psi_\bk$, with $\Psi_\bk^\dagger=(c_{\bk\uparrow}^\dagger,c_{\bk\downarrow}^\dagger, f_{\bk+}^\dagger, f_{\bk-}^\dagger)$ and
\be
H_\bk=\left(\epsilon-2t \sum_\mu \cos k_\mu \right)\sigma_0 \tau_z +2\lambda \tau_y   \sum_\mu \sin k_\mu \sigma_\mu
\end{equation}
where $\sigma_\mu$ ($\mu=x,y,z$) and $\sigma_0=\hat{1}$ act on the (pseudo)spin space, while $\tau_z$, $\tau_y$ (together with $\tau_x$ and $\tau_0$) act on the orbital space.
Applying now the rotation $\tau_i \sigma_j \rightarrow  U'^{-1}\tau_i\sigma_j U'$ with
\be
U'=\left(
\begin{array}{llll}
 1 & 0 &0 &0 \\
0 & 1 &0 &0 \\
0 & 0 &1 &0 \\
0 & 0 &0 &-1
\end{array}\right)=\frac{1}{2}(\sigma_0\tau_0 +\sigma_z\tau_0+\sigma_0\tau_z-\sigma_z\tau_z)
\end{equation}
which just changes the sign of the $|f-\rangle$ state, we get
\begin{eqnarray}
H_\bk&=&\left(\epsilon-2t \sum_\mu \cos k_\mu \right)\sigma_0 \tau_z +\nonumber\\\label{h_sitte}
&&+2\lambda (-\sin k_x  \sigma_y \tau_x + \sin k_y \sigma_x \tau_x + \sin k_z \sigma_0 \tau_y)
\end{eqnarray}
which is the  model of Refs.~\onlinecite{qi_2008,sitte_ti} without inversion-symmetry breaking terms, and $\lambda=-t$.

Similarly, performing a rotation in the orbital space
through $\tau_i \rightarrow  U^{-1}\tau_i U$, with
\be
U=\frac{1}{2}\left(
\begin{array}{ll}
 1-i & 1-i \\
-1-i & 1+i
\end{array}\right)=\frac{1}{2}(\tau_0-i\tau_z-i\tau_x+i\tau_y)
\end{equation}
we get
\be
H_\bk=\left(\epsilon-2t \sum_\mu \cos k_\mu \right)\sigma_0 \tau_x -2\lambda \tau_z  \sum_\mu \sin k_\mu \sigma_\mu
\end{equation}
which is the Hamiltonian studied for example in Ref. \onlinecite{franz_witten}.

Hence, many properties of the mean-field solution of the TKI model apply to the other four-band models as well, with two major differences:
(i) In the common four-band models the two orbitals are often assumed to be roughly equivalent (in fact, fully symmetric within the approximations adopted to obtain Eq.~\ref{h_sitte}). In contrast, in the TKI model the two orbitals are physically very different, i.e., of $c$ and $f$ character.
(ii) Self-consistency leads to an effective model with site-dependent parameters in the presence of impurities and/or surfaces.

Finally, we recall that the TKI model is a genuine many-body model, and, as such, techniques going beyond mean-field, such as dynamical mean-field theory, will reveal further differences with respect to the simple non-interacting four-band model.\cite{assaad_tki_dmft, assaad_temperature}

\section{Spin expectation value of the surface states}\label{app_spin}

To connect to spin-polarized photoemission experiments, we consider a spin-resolved version of the single-particle spectrum and define a generalized spin expectation value $\langle \vec{\sigma} \rangle(\w,\bk, z=1)\equiv \langle \vec{\sigma} \rangle(\omega,\bk)$ on the first layer $z=1$, with $\bk$ being in-plane momentum.
For free particles with $\bk$ being a good quantum number, $\langle \vec{\sigma} \rangle(\omega,\bk)$ is non-zero only if $\w$ matches one of the energy eigenvalues at wavevector $\bk$.
In a clean slab calculation we have eigenvectors $|n\bk \rangle$ and eigenvalues $\epsilon_{n\bk}$ of Eq. \eqref{F_kr}, with $1\le n \le 4 N_z$ , or equivalently, the Green's function
\begin{align}
&\hat G^0(\omega,\bk)_{zsa,z's'a'}\equiv \nonumber\\
&\equiv\langle zsa\bk  |\hat G^0(\omega,\bk)|z's'a'\bk \rangle=\sum_n \frac{A^{n\bk }_{zsa}A^{n\bk *}_{z's'a'}}{\omega-\epsilon_{n\bk }+i \delta},
\end{align}
where $A^{n\bk}_{zsa}\equiv \langle zsa\bk  | n\bk  \rangle$ is the matrix of eigenvectors.

First of all, the non-spin polarized ARPES signal of Figs. \ref{fig_qpi_wti} and \ref{fig_qpi_sti}, panels (a1) to (a4), is obtained through
\begin{eqnarray}
A(\omega, \bk, z=1)&=&-\frac{1}{\pi}\I \sum_{saz=1} \hat G^0(\omega,\bk)_{zsa,zsa}=\nonumber\\
&=&-\frac{1}{\pi}\I\Tr[\hat G^0(\omega,\bk) \hat Z],
\end{eqnarray}
where operator $\hat Z$ is a projector on the subspace with $z=1$:
\be
\hat Z_{zsa,z's'a'}=\delta_{z=z'=1}.
\end{equation}


Now, the total expectation value of the spin at energy $\omega$, plane momentum $\bk$, and on layer $z=1$ is
\begin{eqnarray}
\langle \vec{\sigma} \rangle(\omega,\bk)&=&-\frac{1}{\pi}\I \sum_{n} \frac{\langle n\bk | \vec{\sigma}\hat Z | n\bk  \rangle}{\omega-\epsilon_{n\bk }+i \delta}=\nonumber\\
&=&-\frac{1}{\pi}\I \sum_{n zz' ss' aa'}A^{n\bk *}_{z's'a'}A^{n\bk }_{zsa}\frac{\langle z' s' a'|\vec\sigma \hat Z|z s a\rangle}{\omega-\epsilon_{n\bk }+i \delta}\nonumber\\
&=& -\frac{1}{\pi}\I \Tr [\hat G^0(\omega,\bk) \vec\sigma \hat Z].
\end{eqnarray}
We observe that
\be
\langle z s a|\vec\sigma \hat Z|z' s' a'\rangle=\delta_{z=z'=1}\delta_{aa'}\langle s a|\vec\sigma |s' a\rangle,
\end{equation}
so we only need to compute matrix elements $\langle s a|\vec\sigma |s' a\rangle$, where $a=c/f$, and $s,s'=\uparrow, \downarrow$ if $a=c$, while $s,s'=+,-$ if $a=f$.
To know the expectation value of the spin on $f$ states Eqs. \eqref{+}, \eqref{-}, we trace out the orbital degree of freedom. The non-zero matrix elements are:
\begin{eqnarray}
\langle + | \sigma^x|-\rangle=&\langle - | \sigma^x|+\rangle=&-\cos^2\eta,\\
\langle + | \sigma^y|-\rangle=&\langle - | \sigma^y|+\rangle^*=&i \cos^2\eta,\\
\langle + | \sigma^z|+\rangle=&-\langle - | \sigma^z|-\rangle=&\cos^2\eta-\sin^2\eta.
\end{eqnarray}
For $c$ states, we have trivially:
\begin{eqnarray}
\langle \uparrow | \sigma^x|\downarrow\rangle=&\langle \downarrow | \sigma^x|\uparrow\rangle=&1,\\
\langle \uparrow | \sigma^y|\downarrow\rangle=&\langle \downarrow | \sigma^y|\uparrow\rangle^*=&-i ,\\
\langle \uparrow | \sigma^z|\uparrow\rangle=&-\langle \downarrow | \sigma^z|\downarrow\rangle=&1.
\end{eqnarray}

\begin{figure}[!tbp]
\includegraphics[width=0.49\textwidth]{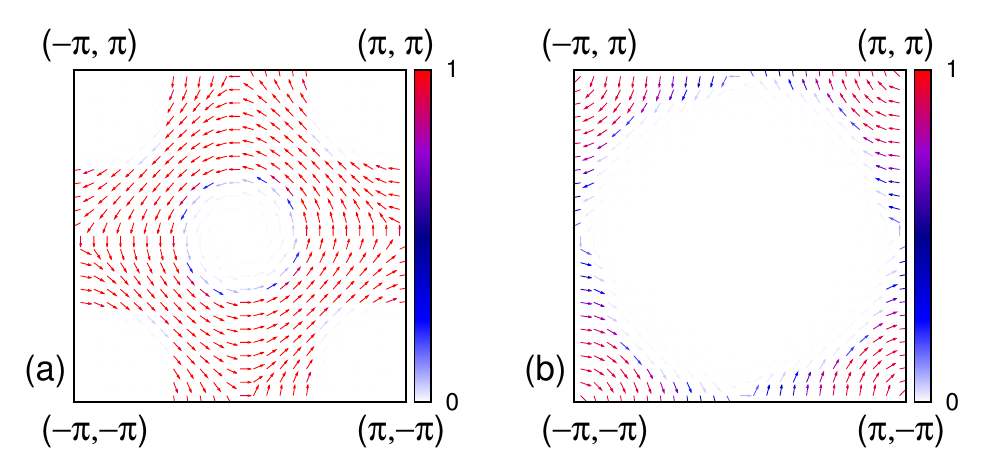}
\caption{Expectation value of the spin in the 2D Brillouin zone, (a) in the WTI phase, integrated from $\omega=0$ to $\omega=0.10$ (relative to the Dirac energy), and (b) in the STI phase integrated from $\omega=0$ to $\omega=0.28$.
The color scale shows the total spin density (arbitrary units), while arrows show the direction.
}\label{fig_spin}
\end{figure}

It turns out that $\langle \sigma^z\rangle$ is negligible close to the Dirac energy,
so the expectation value of the spin in the $f$ shell is simply renormalized by a factor $-\cos^2\eta=-3/7\sim -0.429$, and the total spin points parallel to the surface.

We stress that here operator $\hat Z$ is needed because $\sum_z \langle \vec{\sigma} \rangle(\omega,\bk,z)=0$, since Dirac cones on opposing surfaces have opposite spin expectation values.

Qualitative results of this calculation are shown in Figs. \ref{fig_qpi_wti} and \ref{fig_qpi_sti}, panel (a2), which show where the spin is pointing for momenta belonging to the Fermi surface.
Full results are shown in Fig. \ref{fig_spin}, where $\langle \vec{\sigma} \rangle(\omega,\bk)$ has been integrated over a range of energies close to the Dirac point.
%


\bibliographystyle{apsrev4-1}
\bibliography{tki.bib}

\begin{thebibliography}{10}
\bibitem{tkianderson}%
  \BibitemOpen
  \bibinfo {note} {It has been shown in Ref.~\onlinecite{feng_imp_bulk_ti} that
  the properties of an Anderson impurity model in the bulk of a topological
  Kondo insulator differ from in a trivial insulator.}%
  \bibAnnoteFile{Stop}{tkianderson}%
\end{thebibliography}


\end{document}